\documentclass[]{emulateapj}
\usepackage{times,enumitem}
\usepackage[normalem]{ulem}
\slugcomment{ApJ in press}

\shorttitle{The Compton-thick quasar Mrk~34}
\shortauthors{P. Gandhi et al.}

\def\sax{{\em BeppoSAX}}
\def\xmm{{\em XMM-Newton}}
\def\ch{{\em Chandra}}

\def\swift{{\sl Swift}}

\def\astroh{{\sl Astro-H}}
\def\athena{{\sl Athena}}

\def\hst{{\sl HST}}

\def\rosat{{\em ROSAT}}

\def\suzaku{{\em Suzaku}}

\def\iras{{\em IRAS}}

\def\nustar{{\em NuSTAR}}

\def\p{$\pm$}
\def\ltsim{\mathrel{\hbox{\rlap{\hbox{\lower4pt\hbox{$\sim$}}}\hbox{$<$}}}}
\def\gtsim{\mathrel{\hbox{\rlap{\hbox{\lower4pt\hbox{$\sim$}}}\hbox{$>$}}}}

\def\Msunpyr{M$_{\odot}\,$yr$^{-1}$}
\def\Msun{M$_{\odot}$}
\def\Lsun{L$_{\odot}$}
\def\micron{$\mu$m}
\def\araa{ARA\&A}
\def\aap{A\&A}
\def\nat{Nature}

\def\mnras{MNRAS}
\def\apj{ApJ}
\def\apjs{ApJS}
\def\aj{AJ}
\def\apjl{ApJL}

\def\pasj{PASJ}

\def\nh{$N_{\rm H}$}

\def\xspec{{\sc xspec}}
\def\pexrav{{\sc pexrav}}
\def\pexmon{{\sc pexmon}}

\def\ha{H$\alpha$}
\def\hb{H$\beta$}

\def\oviii{O\,{\sc viii}}
\def\ovii{O\,{\sc vii}}
\def\nex{Ne\,{\sc x}}
\def\oiii{[O\,{\sc iii}]}

\def\nii{[N\,{\sc ii}]}

\def\mbh{$M_{\rm BH}$}
\def\gx339{GX~339--4}
\def\swiftj1753{SWIFT~J1753.5--0129}
\def\xtej1118{XTE~J1118+480}

\def\heasoft{{\sc heasoft}}
\def\caldb{{\sc caldb}}
\def\mytorus{{\sc MYTorus}}
\def\torus{{\sc torus}}
\def\apec{{\sc apec}}

\def\phabs{{\sc phabs}}
\def\apecone{{\sc apec}$_1$}
\def\apectwo{{\sc apec}$_2$}
\def\fscatt{$f_{\rm scatt}$}
\def\loiii{$L_{\rm [O\,{\scriptscriptstyle III}]}$}
\def\lx{$L_{\rm X-ray}$}
\def\ltwoten{$L_{\rm 2-10}$}

\def\lirrest{$L_{\rm 12\mu m}$}

\begin{document}
\title{NuSTAR unveils a Compton-thick Type 2 quasar in Mrk~34}

\author{P. Gandhi$^1$, G.B. Lansbury$^1$, D.M. Alexander$^1$, D. Stern$^2$, P. Ar\'{e}valo$^{3,4}$, D.R. Ballantyne$^5$, M. Balokovi\'{c}$^6$, F.E. Bauer$^{3,7,8}$, S.E. Boggs$^9$, W.N. Brandt$^{10,11}$, M. Brightman$^{12}$, F.E. Christensen$^{13}$, A. Comastri$^{14}$, W.W. Craig$^{13,15}$, A. Del Moro$^1$, M. Elvis$^{16}$, A.C. Fabian$^{17}$, C.J. Hailey$^{18}$, F.A. Harrison$^6$, R.C. Hickox$^{19}$, M. Koss$^{20}$, S.M. LaMassa$^{21}$, B. Luo$^{9,10}$, G.M. Madejski$^{22}$, A.F. Ptak$^{23}$, S. Puccetti$^{24,25}$, S.H. Teng$^{26}$, C.M. Urry$^{21}$, D.J. Walton$^6$, W.W. Zhang$^{23}$}
\affil{$^1$Department of Physics, Durham University, South Road DH1 3LE, UK}
\affil{$^2$Jet Propulsion Laboratory, California Institute of Technology, 4800 Oak Grove Drive, Mail Stop 169-221, Pasadena, CA 91109, USA}
\affil{$^3$Instituto de Astrof\'{\i}sica, Facultad de F\'{i}sica, Pontificia Universidad Catolica de Chile, 306, Santiago 22, Chile}
\affil{$^4$Instituto de F\'isica y Astronom\'ia, Facultad de Ciencias, Universidad de Valpara\'iso, Gran Bretana N 1111, Playa Ancha, Valpara\'iso, Chile}
\affil{$^5$Center for Relativistic Astrophysics, School of Physics, Georgia Institute of Technology, Atlanta, GA 30332, USA}
\affil{$^6$Cahill Center for Astrophysics, 1216 East California, Boulevard, California Institute of Technology, Pasadena, CA 91125, USA}
\affil{$^7$Millennium Institute of Astrophysics}
\affil{$^8$Space Science Institute, 4750 Walnut Street, Suite 205, Boulder, Colorado 80301}
\affil{$^9$Space Sciences Laboratory, University of California, Berkeley, CA 94720, USA}
\affil{$^{10}$Department of Astronomy and Astrophysics, 525 Davey Lab, The Pennsylvania State University, University Park, PA 16802, USA}
\affil{$^{11}$Institute for Gravitation and the Cosmos, The Pennsylvania State University, University Park, PA 16802, USA}
\affil{$^{12}$Max-Planck-Institut f\"{u}r extraterrestrische Physik, Giessenbachstrasse 1, D-85748 Garching bei M\"{u}nchen, Germany}
\affil{$^{13}$DTU Space-National Space Institute, Technical University of Denmark, Elektrovej 327, DK-2800 Lyngby, Denmark}
\affil{$^{14}$INAF Osservatorio Astronomico di Bologna, via Ranzani 1, I-40127, Bologna, Italy}
\affil{$^{15}$Lawrence Livermore National Laboratory, Livermore, CA 94550, USA}
\affil{$^{16}$Harvard-Smithsonian Center for Astrophysics, 60 Garden Street, Cambridge, MA 02138, USA}
\affil{$^{17}$Institute of Astronomy, Madingley Road, Cambridge CB3 0HA, UK}
\affil{$^{18}$Columbia Astrophysics Laboratory, 550 W 120th Street, Columbia University, NY 10027, USA}
\affil{$^{19}$Department of Physics and Astronomy, Dartmouth College, 6127 Wilder Laboratory, Hanover, NH 03755, USA}
\affil{$^{20}$Institute for Astronomy, Department of Physics, ETH Zurich, Wolfgang-Pauli-Strasse 27, CH-8093 Zurich, Switzerland}
\affil{$^{21}$Yale Center for Astronomy and Astrophysics, Physics Department, Yale University, PO Box 208120, New Haven, CT 06520-8120, USA}
\affil{$^{22}$Kavli Institute for Particle Astrophysics and Cosmology, 
Stanford University, 2575 Sand Hill Road M/S 29, Menlo Park, CA 94025, USA}
\affil{$^{23}$X-ray Astrophysics Laboratory, NASA Goddard Space Flight Center, Greenbelt, MD 20771, USA}
\affil{$^{24}$ASDC--ASI, Via del Politecnico, 00133 Roma, Italy}
\affil{$^{25}$INAF--Osservatorio Astronomico di Roma, via Frascati 33, 00040 Monte Porzio Catone (RM), Italy}
\affil{$^{26}$Observational Cosmology Laboratory, NASA Goddard Space Flight Center, Greenbelt, MD 20771, USA}

\label{firstpage}
\begin{abstract}
We present \nustar\ 3--40 keV observations of the optically selected Type 2 quasar (QSO2) SDSS~J1034+6001 or Mrk~34. The high-quality hard X-ray spectrum and archival \xmm\ data can be fitted self-consistently with a reflection-dominated continuum and strong Fe K$\alpha$ fluorescence line with equivalent-width $>1$~keV. Prior X-ray spectral fitting below 10 keV showed the source to be consistent with being obscured by Compton-thin column densities of gas along the line-of-sight, despite evidence for much higher columns from multiwavelength data. \nustar\ now enables a direct measurement of this column, and shows that \nh\ lies in the Compton-thick (CT) regime. The new data also show a high intrinsic 2--10~keV luminosity of $L_{2-10} \sim 10^{44}$~erg~s$^{-1}$, in contrast to previous low-energy X-ray measurements where $L_{\rm 2-10}\ltsim 10^{43}$~erg~s$^{-1}$ (i.e. X-ray selection below 10 keV does not pick up this source as an intrinsically luminous obscured quasar). Both the obscuring column and the intrinsic power are about an order of magnitude (or more) larger than inferred from pre-\nustar\ X-ray spectral fitting. Mrk~34 is thus a \lq gold standard\rq\ CT QSO2 and is the nearest non-merging system in this class, in contrast to the other local CT quasar NGC~6240 which is currently undergoing a major merger coupled with strong star-formation. For typical X-ray bolometric correction factors, the accretion luminosity of Mrk~34 is high enough to potentially power the total infrared luminosity. X-ray spectral fitting also shows that thermal emission related to star-formation is unlikely to drive the observed bright soft component below $\sim$3 keV, favoring photoionization instead.
\end{abstract}

\section{Introduction}

The census of active galactic nuclei (AGN) remains severely incomplete at high gas column densities (\nh). The line-of-sight column \nh(los)\ becomes optically thick to Thomson scattering above $\approx 1.2\times10^{24}$~cm$^{-2}$ for typical cosmic abundances. This is the Compton-thick or \lq CT\rq\ regime. CT sources provide a significant contribution to the cosmic X-ray background (CXB) spectrum, with most CXB models requiring a CT AGN contribution of around 10--25\%\ to the CXB peak flux at $\approx 30$ keV \citep[e.g. ][]{comastri95, g03, gilli07, treister09, draperballantyne10, akylas12, ueda14}. Yet, only $\approx 20$ {\em bona fide} CT Seyferts have been confirmed in the local universe, where \nh(los) can be robustly measured from self-consistent spectral fitting based upon a significant hard X-ray continuum ($>$10 keV) and detection of a strong neutral Fe K$\alpha$ line with equivalent width (EW)\,$\gtsim$\,1 keV \citep{dellaceca08, goulding12}. These features result from suppression of direct photons leading to a dominant contribution from reflection off circumnuclear gas. 

The number density of obscured AGN is even more uncertain at higher luminosities. \lq Type 2 quasars\rq\ (QSO2s) are characterized by luminous narrow emission lines in the optical/infrared and the absence of broad lines, as would be expected if significant circumnuclear obscuration is present. The radio-loud subset of these, or powerful \lq radio galaxies\rq, were the first to be identified \citep[e.g. ][]{mccarthy93, mileydebreuck08}, and X-ray studies have shown many of them to be heavily obscured along the line-of-sight \citep[e.g. ][]{g06_4c3929, tozzi09, wilkes13}. Now, the dominant population of radio-quiet QSO2s is being revealed through homogeneous, wide-area selection both through mid-infrared color selection from {\it Spitzer} and {\it WISE} \citep[e.g. ][]{lacy04, stern05, donley12, stern12, mateos12, assef13}, as well as spectroscopic identification by the Sloan Digital Sky Survey \citep[e.g. ][]{zakamska03, reyes08}. 

Sloan-selected QSO2s, or SDSS-QSO2s, are identified as sources with \loiii~$\gtsim 10^{8.3}$~\Lsun\ for the forbidden \oiii$\lambda$5007\AA\ line and an absence of broad permitted lines, but selection is also based upon redshift-dependent line flux ratios required for removing star-forming galaxies as well as a spectroscopic signal-to-noise selection. X-ray follow-up has shown most SDSS-QSO2s to be consistent with the unified AGN scheme in having significant columns of obscuring gas \citep{vignali04, vignali06, ptak06, jia13}. By using known mid-infrared to X-ray and \oiii\ to X-ray luminosity relations, \citet{vignali10} infer that about half of SDSS-QSO2s are obscured by CT gas columns. This is an important conclusion but is, at present, an {\em indirect} one, being based upon many X-ray non-detections and assumptions of the intrinsic AGN spectral shapes. There is a known degeneracy between the strength of reflection and obscuring column density \citep{g07, treister09} which can only be resolved using high-quality hard X-ray data \citep[e.g. ][]{matt00,ricci11, vasudevan13_localxrb, delmoro14}.

X-ray surveys can potentially address these issues by selecting optically faint and radio-quiet obscured quasars with $L_{2-10\ \rm keV}\gtsim 10^{44}$~erg~s$^{-1}$ and \nh~$> 10^{22}$~cm$^{-2}$ \citep[e.g. ][]{g04, mainieri11, merloni14}. But confirmation of CT X-ray obscuration in distant quasars requires the very deepest pencil-beam surveys, and not many sources have enough counts for detailed spectral characterization \citep{norman02, stern02, feruglio11, comastri11, brightmanueda12, georgantopoulos13}. The study of distant quasars ($z\gtsim 2$) is helped by the redshifting of hard X-ray photons to the energy range below 10 keV where most sensitive X-ray observatories so far have operated. This is not the case for low redshift quasars, leaving significant uncertainties in modeling of their X-ray spectra even though they appear brighter.

As a result of all these issues, the contribution of QSO2 activity to the CXB as well as to AGN growth and evolution remain uncertain. Yet, the latest synthesis models predict that obscured quasars outnumber unobscured ones at high redshift ($z\sim 2$) where the peak of black hole and galaxy growth occur \citep[][]{ueda14}. An accurate census of QSO2s is thus clearly important. 

The {\em Nuclear Spectroscopic Telescope Array} (\nustar, \citealt{nustar}) is the first mission in orbit capable of true imaging at energies of $\sim 3-79$ keV with an angular resolution better than previous hard X-ray observatories by over an order of magnitude. This enables an effective gain of $\gtsim 100$ for direct studies of the broadband hard X-ray continua of a variety of cosmic sources. The CXB peak at $\approx 30$ keV lies within \nustar's operational energy range, and we expect to resolve $\sim 30$~\%\ of its integrated flux \citep{ballantyne11}, as compared to the current $\sim 1$~\%\ level \citep{burlon11, vasudevan13_localxrb}. We may also expect to better elucidate the QSO2 contribution to the overall AGN population. In a first look at the hard X-ray sky with \nustar, \citet{alexander13} found an abundance of mildly obscured quasars selected at $\gtsim 10$ keV ($L_{\rm 10-40\ keV}>10^{44}$~erg~s$^{-1}$, \nh~$\gtsim 10^{22}$~cm$^{-2}$) but none with \nh~$>10^{24}$~cm$^{-2}$, thus limiting the fraction of CT quasars (CT QSOs) above 10 keV to $\ltsim 33$\% over $z=0.5-1.1$. In parallel, a pilot \nustar\ study of three $z\approx 0.5$ SDSS-QSO2 selected as being CT candidates could not confirm the presence of CT obscuration in any, despite the significantly improved hard band sensitivity with respect to prior \ch\ and \xmm\ constraints \citep{lansbury14}, emphasizing the difficulty of studying distant obscured AGN.  

Here, we present \nustar\ observations of the first target in an extended SDSS-QSO2 sample, for which the selection is designed to provide the best {\em direct} constraints possible on the most obscured SDSS-QSO2s. The source, Mrk~34, is a known Type 2 AGN with narrow permitted and forbidden emission lines \citep{heckman81, dahariderobertis88}. A recent detailed multi-component fit to the SDSS nuclear spectrum finds a maximal velocity component with full-width at half-maximum (FWHM) of $\approx 616$~km~s$^{-1}$ to the \ha\ and \hb\ permitted lines and the forbidden \oiii\ and \nii\ doublets \citep{mullaney13}.\footnote{An extra highly broadened component is reported in their fit with FWHM\,=\,11989.1\,km\,s$^{-1}$, but is not significant.} The observed line power \loiii~$=10^{8.8}$~\Lsun\ is comparable to the mean luminosity of radio-quiet Palomar-Green quasars at $z$$\ltsim$0.5 \citep{borosongreen92}. The source is radio-quiet with a small-scale jet of extent 3.7 kpc, bipolar radio morphology ending in two hot spots, and evidence of interaction between the jet and the narrow-line region clouds (\citealt{ulvestad84_v}; \citealt{falcke98}; \citealt{fischer13}). In the infrared, the source has a total power of $L_{\rm 8-1000 \mu m} \approx 2\times 10^{11}$~\Lsun\ \citep{gonzalezdelgado01} and lies in the luminosity regime associated with luminous infrared galaxies (LIRGs). It is also a known luminous H$_2$O megamaser source, which is usually associated with the presence of significant circumnuclear absorption \citep{henkel05, kondratko06b, greenhill08}.

X-ray spectroscopy of Mrk~34 with \xmm\ found a strong Fe K$\alpha$ fluorescence emission line \citep{greenhill08, jia13}. The megamaser and the strong Fe line suggest the presence of Compton-thick material in the source, but this has not been possible to prove with data below 10 keV alone. The \xmm-detected continuum is that of a Compton-thin (\nh~$<10^{24}$~cm$^{-2}$) AGN when fitted with an absorbed power-law. In the \nustar\ data presented herein, we detect the source to $\sim 40$ keV. The high quality spectra unambiguously show, for the first time in a SDSS-selected QSO2, evidence for a reflection-dominated continuum requiring CT absorption along the line-of-sight. This demonstrates the gain that \nustar\ is providing for obscured AGN studies. 

The source redshift is $z$\,=\,0.051, giving a luminosity distance of 236~Mpc for a flat cosmology with $H_{\rm 0}=67.3$~km~s$^{-1}$~Mpc$^{-1}$ and $\Omega_\Lambda=0.685$ \citep{planckcosmology}. All fit uncertainties are quoted at 90\%\ confidence, unless stated otherwise. An outline of this paper is as follows. Section\,2 describes our target selection strategy and section\,3 details of X-ray observations and reduction. The spectral fitting procedures and results from the \nustar\ data alone, and when combined with \xmm, are described in sections\,4 and 5, respectively. In section 6, we discuss the intrinsic source properties, and how Mrk\,34 fits in the context of bona fide CT AGN and the QSO2s population in general. The paper concludes with a summary in section\,7.

\section{Target selection}
\label{sec:sampleselection}

Mrk 34 was chosen as a promising target for \nustar\ from the samples of SDSS-QSO2s that have prior X-ray follow-up observations \citep{vignali06, vignali10, jia13}. By fitting \xmm\ data of the source, \citet{jia13} found an observed (absorbed) \ltwoten~$=9\times 10^{41}$~erg~s$^{-1}$ and \nh~$=2.63_{-2.63}^{+4.21}\times 10^{23}$~cm$^{-2}$ (consistent with an upper limit of \nh\,$< 6.84\times 10^{23}$\,cm$^{-2}$). In addition, Mrk 34 showed the following characteristics associated with heavy or CT X-ray obscuration: 

\begin{enumerate}[label=\bfseries \arabic*)]

\item As compared to the observed power in the \oiii\ emission line (\loiii~$=10^{8.8}$~\Lsun; \citealt{reyes08}), the observed 2--10 keV power is low: \ltwoten/\loiii=0.4.\footnote{No luminosity uncertainties are stated by \citet{jia13}. From our spectral fit, we estimate a 25\% uncertainty on \ltwoten, which will dominate the error in the \ltwoten/\loiii\ ratio.} This ratio places Mrk~34 at about 200$\times$ lower in \ltwoten\ than the local Type 1 AGN correlation between $L_{2-10}$ and \loiii\ \citep{mulchaey94, netzer06, panessa06};

\item The source also shows a low observed \ltwoten/\lirrest\ ratio that places it $\approx 100 \times$ below the correlation (in terms of \ltwoten) presented in \citet{g09_mirxray}. The infrared luminosity is \lirrest~=~$2(\pm 0.04)\times 10^{44}$~erg~s$^{-1}$, measured as $\lambda L_\lambda$, from linear interpolation of multi-band all-sky catalog data produced by the {\em WISE} mission to a rest-frame wavelength of 12~\micron\ \citep{wise};

\item Finally, the \xmm\ data show a strong Fe K$\alpha$ equivalent width (EW$_{\rm K\alpha}$)~$=1.6_{-0.8}^{+0.9}$ keV \citep[][\citealt{greenhill08}]{jia13}. 

\end{enumerate}

\noindent
\citet{jia13} determined an intrinsic X-ray luminosity $L_{\rm 2-10, in} = 2\times 10^{42}$~erg~s$^{-1}$ after correcting for the obscuring column of \nh~$\approx 3\times 10^{23}$~cm$^{-2}$ measured in the \xmm\ data below $\sim 10$ keV. Using physical reflection model fits to the same data, \citet{lamassa14} constrained \nh~$>4\times10^{23}$~cm$^{-2}$ implying a factor of a few larger luminosity, $L_{\rm 2-10, in}=9_{-4}^{+6}\times 10^{42}$~erg~s$^{-1}$. By contrast, using either \loiii\ or \lirrest\ (points 1 and 2 above) as an indirect proxy of the intrinsic power would imply $L_{\rm 2-10, in}\sim 10^{44}$~erg~s$^{-1}$, about 10 times higher still. In addition, the value of EW$_{\rm K\alpha}\gtsim 1$ keV above classifies Mrk~34 as a CT candidate, which would also imply the need for strong corrections to the observed luminosity. This can be tested using higher energy X-ray data. 

The source is not detected by the \swift/BAT all-sky survey at the nominal 70-month survey 4.8-$\sigma$ sensitivity flux limit of $F_{14-195}$=1.3$\times$10$^{-11}$~erg~s$^{-1}$~cm$^{-2}$ \citep{baumgartner13}. But direct examination of the BAT maps reveals a 3.7-$\sigma$ excess at the position of Mrk~34 (for details on the maps and the procedure, see \citealt{koss13}). Extrapolation of the observed \xmm\ continuum to hard X-rays using typical reflection models (described below) also implied a good detection probability with \nustar\ in modest exposure times. Therefore, the object was chosen as a promising \nustar\ target. 

\section{Observations}

\subsection{\nustar}

Mrk~34 was observed on UT2013-09-19 (ObsID 60001134002) for an on-source time of 25.7 ks. The data were processed with the \nustar\ Data Analysis Software ({\sc nustardas}) v.1.3.0 within \heasoft\ v.6.15.1. With the \nustar\ UT2013-05-09 \caldb\ release, calibrated and cleaned event files were produced using the {\sc nustardas} task {\tt nupipeline} with standard filter flags. Passages of the satellite through the South Atlantic Anomaly were filtered out and the standard depth cut used to help reduce instrumental background. Spectra and response files were extracted using the {\tt nuproducts} task. The net exposure time was approximately 23.9~ks. 

A 45\arcsec--radius aperture was used for extracting source counts. Background counts were extracted from a neighboring, off-target circular aperture 100\arcsec\ in radius, free of any other sources. Using a source-centered annular region to define the background gave consistent results. 

\subsection{\xmm}

Archival \xmm\ observations (ObsID 0306050701) obtained through a medium filter in full-window mode from UT2005-04-04 were reduced and analyzed using standard procedures within {\sc sas} v13.0.0.\footnote{{\tt http://xmm.esa.int/sas}} Spectra were extracted within a 15\arcsec--radius aperture. The spectra and responses from the two EPIC {\sc mos} detectors were combined using the {\tt addascaspec} script as they were very similar. The net good exposure time is 8.8 ks ({\sc pn}) and 22.9 ks (combined {\sc mos}). We note that there are no other bright sources within $\sim$1\farcm5 of the target position in the \xmm\ (and \nustar) images. More details on these \xmm\ data can be found in \citet{jia13} and \citet{lamassa14}. 

\section{X-ray spectral fitting}

\begin{figure}
  \begin{center}
  \includegraphics[angle=270,width=8.5cm]{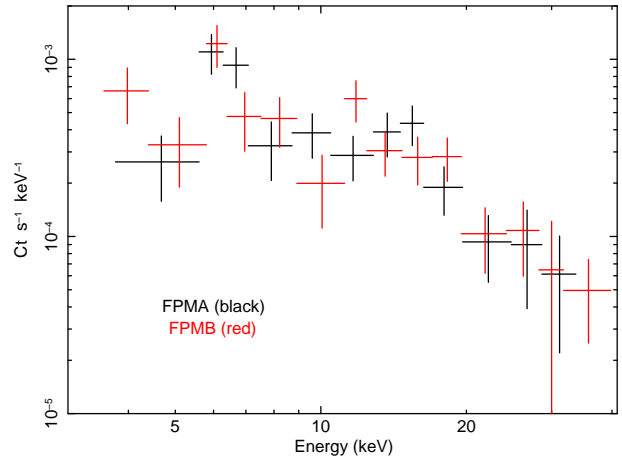}
  \caption{Background-subtracted \nustar\ spectra of Mrk 34 for the two FPMs. 
    \label{fig:basicplot}}
  \end{center}
\end{figure}

We began with an examination of the \nustar\ data alone. Mrk~34 is well detected in both focal plane modules (FPMs), with net count rates of $7.5(\pm 0.7)\times10^{-3}$~ct~s$^{-1}$ in FPMA and $7.9(\pm 0.8)\times10^{-3}$~ct~s$^{-1}$ in FPMB, respectively, over the energy range $\approx 3-40$ keV. The extracted spectra are shown in Fig.~\ref{fig:basicplot}. Spectral analysis was carried out over this energy range using \xspec\ v.12.8.1 \citep{xspec}. All spectra have been grouped to a minimum of 20 counts per bin.

A prominent excess of counts just above 6 keV suggests the presence of Fe K$\alpha$ emission. Parametrizing the data with a power-law (PL) continuum and a Gaussian line with fixed rest-frame line centroid energy $E=6.40$ keV and redshift $z=0.051$ returns a photon index $\Gamma=0.15_{-0.16}^{+0.45}$ (where photon flux density $N_{\rm E}\propto  E^{-\Gamma}$), and EW$_{\rm K\alpha}=1.9_{-0.8}^{+2.3}$~keV with a fit statistic of $\chi^2$\,=\,29.9 for 23 degrees of freedom (dof). The confidence ranges on EW$_{\rm K\alpha}$ are determined by drawing an ensemble of 10,000 parameter values from the fit. Letting the line centroid float freely returns a rest-frame energy $E=6.6_{-0.3}^{+0.1}$~keV, showing that this is consistent with neutral Fe K$\alpha$ (although weaker Fe K$\beta$ or higher ionization lines are likely to be contributing). The observed fluxes are $F_{2-10}=1.9_{-1.0}^{+0.1}\times10^{-13}$~erg~s$^{-1}$~cm$^{-2}$ and $F_{10-30}=1.1_{-0.5}^{+0.1}\times10^{-12}$~erg~s$^{-1}$~cm$^{-2}$, respectively. The very flat $\Gamma$ and large EW$_{\rm K\alpha}$ are characteristic signatures of reflection from optically thick cold gas. 

Over the common energy range of $\sim$3--10 keV, \xmm\ gives consistent results with \nustar\ on spectral shape, fluorescence strength and continuum normalization: $\Gamma=0.17\pm 0.51$, EW$_{\rm K\alpha}=2.1_{-1.0}^{+3.2}$~keV, $F_{2-10}=1.7_{-1.0}^{+0.2}\times 10^{-13}$~erg~s$^{-1}$~cm$^{-2}$. This consistency implies that there has been no significant variability between the two observations. We therefore proceeded to fit more physical reflection models to the \nustar\ and \xmm\ data combined. We used two physically motivated and self-consistent reflection models for the hard X-ray regime, in addition to including other common spectral components at soft energies. These are described below. 

\subsection{\mytorus\ : model M}

The \mytorus\ model simulates a toroidal absorber geometry with a covering factor of 0.5 (a fixed half-opening angle $\theta_{\rm tor}=60$~deg) centered on a continuum source \citep{mytorus}. It self-consistently includes i) distortion of the zeroth-order transmitted component (which, in our case, is a PL) due to photoelectric absorption and Compton scattering, ii) Compton-scattering off the torus, and iii) associated fluorescence line emission (neutral Fe K$\alpha$ at 6.4 keV and K$\beta$ at 7.06 keV) and the Compton shoulder to the fluorescence lines. We used \mytorus\ mainly in the standard (\lq coupled\rq) mode where the normalizations of these three components are tied to the intrinsic continuum, effectively coupling \nh(los) to the torus inclination angle. Table grids with an equatorial column density (\nh(eq)) of up to 10$^{25}$ cm$^{-2}$ are available and defined between 0.5--500 keV. This model is referred to as \lq model M\rq.

\mytorus\ also allows decoupling the column density from geometry, and \nh(los) from that responsible for the scattered continuum (\nh(scatt)). Such a \lq decoupled\rq\ mode fit is briefly described in \S~\ref{sec:intrinsicpower}.

\subsection{\torus\ : model T}

The {\sc torus} model of \citet[][]{brightmannandra11} allows changing geometries through variable $\theta_{\rm tor}$, but \nh(los) through the torus is defined such that it is independent of inclination ($\theta_{\rm inc}$). Table models with \nh\ up to 10$^{26}$ cm$^{-2}$, or about ten times more than allowed in \mytorus, are publicly available. {\sc torus} also includes Compton-scattering and Fe K$\alpha$ fluorescence. It is defined between 0.1--320 keV, which enables us to extend the fit to lower energies as compared to \mytorus. This model is referred to as \lq model T\rq.

\subsection{Other spectral components}

The low-energy ($\ltsim$2 keV) spectral shape as seen with \xmm\ is much softer than the continuum shape at higher energies \citep{jia13, lamassa14}. This was parametrized as hot gas plasma emission in the host galaxy with \apec\ \citep{apec} and included in both models M and T. Two such \apec\ components were found to be necessary for both models. Additional emission from an optically thin medium on larger scales in the host galaxy which scatters AGN emission into the line-of-sight may be present, as could emission from X-ray binaries. Such emission was simulated using a single PL (of photon-index $\Gamma$ tied to that of the AGN) with a scattering fraction $f_{\rm scatt}$ relative to the intrinsic PL, and obscured by an additional gas column (\nh(host)) that is independent of the column associated with the torus. Such components are often used to describe the soft emission observed in obscured AGN (e.g. \citealt{done03}), but in the absence of high spectral resolution soft X-ray data, these are only meant as a simple prescription to describe the spectral shape in this regime. We will discuss whether such components are viable in section~\ref{sec:soft}, together with other physical models of the soft emission.

All abundances for the torus models and the thermal components are fixed at Solar.

\phabs\ absorption through a fixed low Galactic column \nh(Gal)=6.8$\times$10$^{19}$ cm$^{-2}$ of cold gas, based upon H~{\sc i} measurements along the line-of-sight \citep{dickeylongman90}, was also included in all spectral fits. Finally, a cross-calibration constant between the two missions was included as a free parameter. Cross-calibration of the \xmm\ EPIC MOS and pn cameras has shown that these instruments are in very good agreement with each other when fitting over their full energy range \citep{kirsch04}, so we fixed their relative cross-normalization to 1.\\

\noindent
The final models have the following notations in \xspec, with explanatory mappings in square brackets:

\begin{eqnarray}
\textsc{model\ M\ =\  const $\times$ phabs[} \mapsto N_{\textrm {\tiny H}}^{\textrm {\tiny Gal}}{\textsc ]} \times \textsc{zphabs[} \mapsto N_{\textrm {\tiny H}}^{\textrm {\tiny host}}{\textsc ]} \times \nonumber\\
(\ \textrm{\textsc{apec}} (\times 2) + \nonumber\\
\textsc{pow} * \textsc{etable} \{{\tt mytorus\_Ezero\_v00.fits}\} + \nonumber\\
\textsc{atable}\{{\tt mytorus\_scatteredH500\_v00.fits}\} + \nonumber\\
\textsc{atable}\{{\tt mytl\_V000010nEp000H500\_v00.fits}\} + \nonumber\\
\textsc{const [} \mapsto f_{\rm scatt} {\textsc ]}\times\textsc{pow}\ ), \nonumber
\end{eqnarray}

\noindent
and

\begin{eqnarray}
\textsc{model\ T\ =\  const $\times$ phabs[} \mapsto N_{\textrm {\tiny H}}^{\textrm {\tiny Gal}}{\textsc ]} \times \textsc{zphabs[} \mapsto N_{\textrm {\tiny H}}^{\textrm {\tiny host}}{\textsc ]} \times \nonumber\\
(\ \textrm{\textsc{apec}} (\times 2) + \nonumber\\
\textsc{pow} * \textsc{atable}\{{\tt torus1006.fits}\} + \nonumber\\
\textsc{const [} \mapsto f_{\rm scatt} {\textsc ]}\times\textsc{pow}\ ). \nonumber
\end{eqnarray}

\section{Results from combined NuSTAR and XMM-Newton fits}
\label{sec:results}

We first fitted the broadband \nustar\ and \xmm\ data using the coupled \mytorus\ model M. A fit was found with reflection {\em dominating} at all energies above $\sim 3$ keV. This is shown in Fig.~\ref{fig:mainfit} and the best-fit parameters are listed in Table~\ref{tab:xspec}. The fit returned an equatorial column $N_{\rm H}{\rm (eq)}\approx 10 ^{25}$~cm$^{-2}$. The inclination angle lies between $\approx 60-75$~deg, constrained at the lower end by the opening angle of the torus. \nh(eq) lies well within the CT regime. In the coupled \mytorus\ model, the line-of-sight column is tied to \nh(eq) and $\theta_{\rm inc}$ from the model geometry, and \nh(los)~$\approx 2.5\times 10^{24}$~cm$^{-2}$ for the best-fit value of $\theta_{\rm inc}\approx 61$~deg. The upper model threshold of 10$^{25}$ cm$^{-2}$ is permitted by the fit, i.e. \nh\ is unconstrained at the upper end. 

The absorbed luminosity over 2--10 keV is $L_{2-10}=1.2\times 10^{42}$\,erg\,s$^{-1}$, and over a broader energy range of 0.5--30 keV covering both \nustar\ and \xmm\ is $L_{0.5-30}=8.8\times 10^{42}$\,erg\,s$^{-1}$. Correcting for absorption implies $L_{\rm 2-10,in}~\approx~6\times~10^{43}$~erg~s$^{-1}$ and $L_{\rm 0.5-30,in}~\approx~1.5~\times~10^{44}$~erg~s$^{-1}$, respectively. The observed EW$_{\rm K\alpha} \approx 1.2$ keV, as expected for a reflection-dominated spectrum.\footnote{Measured by numerical integration of the best-fit Fe K$\alpha$ line plus Compton shoulder model flux, and dividing by the total fitted \mytorus\ continuum interpolated to 6.4 keV rest-frame.} The fitted $\Gamma \approx 2.2$ is somewhat steep compared to the mean $\langle\Gamma\rangle\sim 1.9$ measured in high-quality AGN data \citep[e.g. ][]{nandra97, mateos05_wide, piconcelli05}, but not extraordinarily so within the allowed confidence range. 

The softest energies can be parametrized using two \apec\ components and optically thin scattering. The temperatures ($kT$) of the \apec\ components are $\approx$0.1 and 1 keV, respectively. Removing either \apec\ component results in a $\Delta$$\chi^2$ of at least +50 for two additional degrees of freedom, implying that both are required. The luminosity associated with the two \apec\ components is $L_{0.5-2}^{\rm APEC} \approx 4\times10^{41}$~erg~s$^{-1}$. The scattering normalization $f_{\rm scatt} \sim 0.3\%$, which gives a luminosity $L_{2-10}^{\rm scatt} \sim 10^{41}$~erg~s$^{-1}$. The \xmm--to--\nustar\ cross-calibration constant is lower than (but consistent with)~1.

Notice that the best-fit \nh(eq) and $\theta_{\rm inc}$ values are at the upper and lower extremes, respectively, allowed in \mytorus\ for an obscured line-of-sight intersecting the torus. This suggests that there is some tension in the fit between the conflicting need to strongly suppress the intrinsic continuum at all energies (i.e. a large \nh) and the need for an unabsorbed reflection-dominated continuum and strong Fe K$\alpha$ line below 10 keV (which pushes $\theta_{\rm inc}$ down). This is not easy to simulate within \mytorus, because of its geometrically constrained configuration. Removing one degree of freedom by fixing $\theta_{\rm inc}$ to 75 deg (at the high end of the allowed confidence range in the above fit) also yields an acceptable solution with \nh(los)~$=3\times 10^{24}$~cm$^{-2}$ and $L_{\rm 2-10,in}=1.7\times 10^{44}$~erg~s$^{-1}$. 

Model T gives the opportunity to examine systematically different torus model assumptions and geometry, in particular in its allowance of varying opening torus angles. We found that it was not possible to constrain both $\theta_{\rm inc}$ and $\theta_{\rm tor}$ simultaneously. We thus decided to fix $\theta_{\rm inc}=87$ deg, corresponding to an edge-on inclination. The strong megamaser in Mrk~34 provides some justification for this choice, as most megamasers are associated with edge-on inclinations \citep{kuo11}. This choice also allows examination of the effect of a larger $\theta_{\rm inc}$ than preferred by \mytorus. In addition, the fit for $\theta_{\rm tor}$ is then no longer restricted by $\theta_{\rm inc}$, as the model allows for $\theta_{\rm tor}$ values up to 84 deg \citep{brightmannandra11}. 

Model T yields an excellent fit, with a flatter $\Gamma \approx 1.7$ (Table~\ref{tab:xspec}, Fig.~\ref{fig:mainfit}). The best-fit \nh(los) lies above 10$^{25}$~cm$^{-2}$ with the lower 90~\%\ confidence interval allowing \nh(los)~$> 3.5\times 10^{24}$~cm$^{-2}$. This is again within the CT regime and unconstrained at the high end, this time up to $10^{26}$~cm$^{-2}$. In this case, EW$_{\rm K\alpha} \approx 1.4$ keV. A broad range of torus opening angles is allowed, $27\ltsim \theta_{\rm tor}\ltsim 78$~deg. Small $\theta_{\rm tor}$ values are equivalent to large torus solid angles, which can produce stronger reflection and fluorescence components. On the other hand, very thin tori ($\theta_{\rm tor}> 78$~deg in our fits) do not produce enough reflection. Whereas model M requires some host absorption (\nh(host)~$\approx 6 \times 10^{21}$~cm$^{-2}$), this is partly driven by the fitted steep $\Gamma$ value and is not needed for model T. 

Finally, we note that using the \pexrav\ \citep{pexrav} or \pexmon\ \citep{pexmon} reflection models as alternatives to the torus models above give broadly similar results in terms of the requirement of a reflection-dominated continuum. In fact, reflection-only models ($R$ value of less than 0) yield fully acceptable fits (with reduced $\chi^2 \approx 1$) without the need for any direct component. These models are not described in detail here because they simulate reflection off a slab geometry and assume an infinite optical depth, neither of which are likely to represent the torus. There is also no strong constraint on the intrinsic luminosity in such a scenario because the reflector and obscurer are disjoint and the solid angle of the reflecting surface visible to us is unknown. 

\section{Discussion}
\label{sec:discussion}

Mrk~34 was selected for \nustar\ observation because previous low-energy X-ray follow-up provided indirect evidence of high or CT obscuration, including low $L_{2-10}$/\loiii\ and low $L_{2-10}$/\lirrest\ ratios relative to Type~1 AGN, as well as an EW$_{\rm K\alpha} > 1$ keV. In terms of its optical spectral properties, in particular its narrow-line luminosity (\loiii~$=10^{8.8}$~\Lsun), Mrk 34 is quite representative of the large sample of 887 SDSS-QSO2s compiled by \citet{reyes08} whose distribution has an average $\langle$\loiii$\rangle =10^{8.7}$~\Lsun\ (with a 1-$\sigma$ scatter of 0.4 dex). With \nustar, we find {\em direct} evidence of CT obscuration in the form of a flat hard X-ray (3--40 keV) continuum which can be modeled as arising from reflection off optically thick circumnuclear gas self-consistently with the strong neutral Fe line. 

Although the source was selected from optical spectroscopy, the derived intrinsic X-ray power also places Mrk~34 in (or very near) the regime associated with obscured X-ray quasars ($L_{\rm 2-10,in} \gtsim 10^{44}$~erg~s$^{-1}$). Whereas local CT Seyfert 2s have been studied at lower X-ray luminosities for many years, this is the first time that such a robust measurement of the intrinsic X-ray power is possible for a powerful SDSS-selected QSO2. Mrk 34 is thus a \lq gold standard\rq\ CT QSO2, satisfying all the above characteristics typically associated with this class.

\subsection{On the intrinsic power and obscuring column density}
\label{sec:intrinsicpower}

The two broadband \nustar+\xmm\ model fits in Table~\ref{tab:xspec} require \nh(los)~$\gtsim 2\times 10^{24}$~cm$^{-2}$ although much higher values of \nh\ are allowed. The photon indices of the two fits also allow for some variation in the intrinsic PL slope. Despite these uncertainties, it is noteworthy that the corresponding absorption-corrected X-ray luminosities of the two model fits differ only by a factor of 2, at $L_{\rm 2-10,in}\sim {\rm (0.6-1.2)}\times 10^{44}$~erg~s$^{-1}$. This is a consequence of the fact that the observed hard X-ray spectrum is dominated by the reflection component at all energies, so the modeled intrinsic luminosity is mainly dependent upon the observed flux and the covering factor of the reflector and is less sensitive to the reflecting gas column or the intrinsic PL shape that produces the reflection spectrum. In other words, the model fits likely provide a good estimate of the intrinsic power of the AGN, but only a lower limit on the torus column density. 

A systematically different obscuring geometry could yield a different result, of course, and this may be investigated using the \lq decoupled\rq\ mode of \mytorus\ \citep{mytorus}. With differing normalizations and/or column densities for the transmitted and scattered components, this mode can be interpreted as an approximate parametrization of alternate geometries such as a patchy torus or of differing elemental abundances. But it is difficult to constrain such decoupled models for Mrk~34 because the intrinsic continuum is not obviously visible at any energy, leading to a large degeneracy on the reflection fraction. In fact, including separate edge-on and face-on scattering/fluorescence components with a free multiplicative scaling between them in the \mytorus\ fit (as recommended by \citealt{yaqoob12}) leads to the scattering component completely dominating over the intrinsic PL by large factors ($\sim$100) suggesting much higher values of \nh(los) than the lower confidence ranges that we presently constrain.

\begin{figure}
  \begin{center}
  \includegraphics[width=8.5cm]{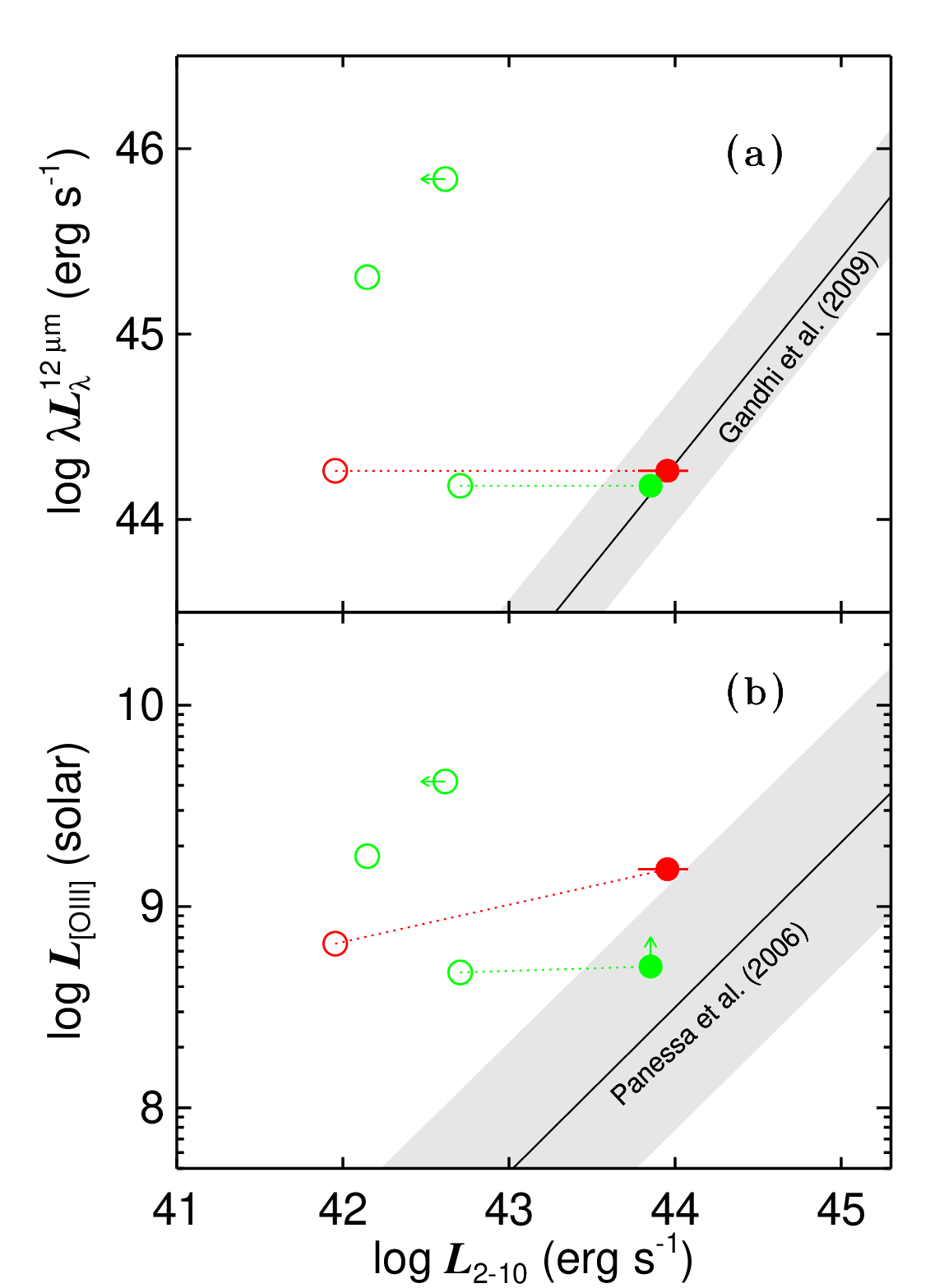}
  \vspace*{0.5cm}
  \caption{Multiwavelength correlations: {\bf (a)} \lirrest\ vs. $L_{2-10}$, and {\bf (b)} \loiii\ vs. $L_{2-10}$, for Mrk~34 (red) and the three other \nustar--observed QSO2s from \citet[][green]{lansbury14}. Open symbols denote observed luminosities. Filled symbols denote intrinsic values for the two \nustar-detected sources (Mrk~34 in red and SDSS~J0011+0056 in green). These are corrected for obscuration and reddening as described in the text. For Mrk~34, the plotted X-ray luminosity is the mean of the two model best-fit values of $L_{\rm 2-10,in}$ in Table~\ref{tab:xspec}, with the uncertainty denoting the range between them. The upward-pointing arrow for SDSS~J0011+0056 denotes the fact that the Balmer decrement and narrow-line-region reddening correction estimate are not available in this case. The shaded zones denote the 1-$\sigma$ relation scatters of $\approx$0.3 and 0.6 dex, respectively. 
    \label{fig:correlations}}
  \end{center}
\end{figure}

Additional checks on the intrinsic power can be provided by indirect multiwavelength relations. These are shown in Fig.~\ref{fig:correlations}, the top panel of which plots the \lirrest\ vs. \ltwoten\ relation for Mrk~34 along with the three other SDSS-QSO2s observed by \nustar\ so far \citep{lansbury14}. The latter objects were selected from sources at $z\sim 0.4-0.5$ with \ltwoten/\loiii~$<2.5$ and sample a range of about one dex in \loiii. The intrinsic $L_{2-10}$ for Mrk~34 matches well with the infrared:X-ray luminosity relation by \citet{g09_mirxray}, lying within the 1-$\sigma$ relation scatter. The same holds true when using other relations published in \citet{g09_mirxray} and \citet{asmus11} which effectively include varying levels of host galaxy contamination to the mid-infrared. Although high angular resolution infrared imaging (e.g. \citealt{asmus14}) is not available for Mrk~34, we note that the {\em WISE} mid-infrared color of the source from the all-sky catalog is $W1-W2=1.18\pm 0.03$, placing the source comfortably inside the color zone identified by \citet{stern12} where the mid-infrared flux is likely to be AGN-dominated. In other words, the close match to the correlation in the top panel of Fig.~\ref{fig:correlations} is entirely attributable to the AGN power alone. 

The same figure also shows the \loiii\ vs. \ltwoten\ plane. Mrk~34 lies at the threshold of the 1-$\sigma$ scatter of the relation defined by \citet{panessa06}, and the source position is approximately similar with respect to the mean and scatter defined by other relations (e.g. \citealt{netzer06}). {\em Intrinsic} \oiii\ luminosities are also shown for the two \nustar-detected sources, and are corrected for Galactic reddening \citep{schlafly11}. For Mrk~34, we were able to additionally correct for reddening in the narrow-line-region of the host galaxy, following the procedure of \citet{bassani99}. For this, we measured the Balmer decrement from the \ha\ and \hb\ narrow line fluxes provided by the SDSS {\tt spZline} pipeline output \citep{bolton12}. A ratio of $F_{\rm H\alpha}/F_{\rm H\beta}=3.98\pm 0.03$ is measured\footnote{It is worth noting that this Balmer decrement is much milder than the value of $F_{\rm H\alpha}/F_{\rm H\beta}$=10.47 according to \citet{dahariderobertis88}, which may be a result of differing slit positioning and setup. We use the more recent SDSS measurements here.}, which translates into an \oiii\ dereddening factor of 2.3. This correction is not possible for the sources from \citet{lansbury14} because their higher redshift means that \ha\ lies beyond the SDSS wavelength range. According to the bottom panel of Fig.~\ref{fig:correlations}, there is a mild suggestion that Mrk~34 may have a higher \ltwoten\ still, but it is difficult to provide a precise cross-check given the large scatter associated with most \loiii:\ltwoten\ correlations. 

Finally, with an isotropic megamaser power $L_{\rm H_2O}$$\sim$10$^3$ \Lsun\ \citep{henkel05}, Mrk~34 now also lies within the scatter of the possible relationship between $L_{\rm H_2O}$ and $L_{2-10}$ proposed by \citet{kondratko06b}. These authors identified two maser complexes in Mrk~34 consistent with symmetric blue- and redshifted high-velocity ($\sim$500 km s$^{-1}$) emission. This is consistent with maser excitation resulting from X-ray irradiation of accretion disk gas. Higher intrinsic X-ray luminosities would push the source beyond the scatter of the proposed $L_{\rm H_2O}$:$L_{2-10}$ relation. 
 
In short, these multiwavelength comparisons suggest that our X-ray spectral analysis reliably captures the intrinsic X-ray power of Mrk~34. A compilation of luminosities from our \nustar\ modeling and over the various wavelength bands discussed above is presented in Table~\ref{tab:lum}. 

\subsection{On the bolometric luminosity and Eddington ratio}

When compared to the total infrared power of $L_{8-1000} \approx 2\times 10^{11}$~\Lsun\ $=8\times 10^{44}$~erg~s$^{-1}$ as probed by \iras, $L_{\rm 2-10,in}$/$L_{8-1000}=0.08-0.15$. For any typical AGN bolometric correction factor $L_{\rm Bol}$/$L_{\rm 2-10}\approx 10-30$ \citep[e.g. ][]{elvis94, vasudevan10}, the AGN easily has enough power to drive the bulk of the infrared emission. LIRGs (with $10^{11}$~\Lsun~$< L_{8-1000} < 10^{12}$~\Lsun) generally show a much lower fractional AGN contribution to the infrared \citep[e.g. ][]{alonsoherrero12}, and Mrk~34 appears to be more similar to Palomar-Green QSOs in this respect \citep{veilleux09}. The high current luminosity of Mrk\,34 may simply represent an upward fluctuation in an otherwise more modest accretion history \citep[e.g. ][]{hickox14}, although a significant ongoing gas accretion rate is required in order to drive the observed power: $\dot{M}=L_{\rm Bol}/\eta c^2 \approx 0.2-0.3$~\Msun~yr$^{-1}$ (assuming an X-ray bolometric correction of 15 and accretion efficiency $\eta=0.1$).

There is no secure measurement of the supermassive black hole mass (\mbh) for Mrk~34 as yet. Although a megamaser has been detected, there is no corresponding spatially resolved map so its inner radius is unknown. However, if we assume that the megamaser disk in Mrk~34 has an inner radius of $\sim 0.1-0.5$~pc as seen in other nearby sources \citep[e.g. ][]{kuo11}, the observed 500~km~s$^{-1}$ maser velocity \citep{kondratko06b} would imply \mbh~$\sim (0.6-3)\times 10^7$~\Msun. Extrapolating the galaxy stellar velocity dispersion ($\sigma$) from the \oiii\ emission line width and using the \mbh--$\sigma$ relation from \citet[][]{tremaine02}, \citet{wang07_feedback} instead estimate \mbh~$=10^{7.96}$~\Msun. On the other hand, \citet{oh11} measure $\sigma$\,=\,$123$\,\p\,5\,km\,s$^{-1}$ from their analysis of the SDSS spectra, and using this velocity dispersion together with the updated \mbh--$\sigma$ relation from \citet{mcconnellma13} would imply \mbh~$=10^{7.12\pm0.38}$~\Msun. 

If we assume that a range of \mbh~$=10^{7-8}$~\Msun\ reasonably encompasses the current uncertainty on \mbh, and combine the intrinsic X-ray power that we measure with typical AGN bolometric correction factors quoted above, we find that the AGN is radiating at an Eddington fraction of $\approx 0.05-2.5$. 

Better data are clearly needed for more precise measurements of \mbh. A very long baseline interferometry map of the megamaser would resolve the inner disk radius, and hence provide a measure of \mbh. At the distance of Mrk~34, a physical scale of 0.5~pc corresponds to an angular size of $\approx 0.48$~mas, which is within range of the synthesized beam sizes currently available \citep{kuo11}. Alternatively, high-quality near-infrared imaging with the {\em Hubble Space Telescope} (\hst) or the future {\em James Webb Space Telescope} could provide a complementary measurement of \mbh\ through accurate measurement of the bulge mass and the known correlation between the two quantities \citep[e.g. ][]{marconihunt03}.

\begin{table}
\begin{center}
 \caption{Mrk\,34 luminosities\label{tab:lum}}
 \begin{tabular}{lr}
  \hline
Band and quantity     &  10$^{43}$ erg\,s$^{-1}$\\
  \hline
  $L_{\rm 0.5-2, in}$ & 0.02 \\
  $L_{2-10}$ (absorbed) & 0.12 \\
  $L_{0.5-30}$ (absorbed) & 0.88 \\
  $L_{10-40}$ (absorbed) & 1.02 \\
  $L_{\rm 2-10,in}^\dag$ & 6--12 \\
  $L_{\rm 0.5-30,in}^\dag$ & 15--31 \\
  $L_{\rm 10-40,in}^\dag$ & 4--15 \\
  \loiii\ (reddened) & 0.24 \\
  \loiii\ (dereddened) & 0.56 \\
  $\lambda L_{\lambda}$(12\,$\mu$m) & 20 \\
  $L_{\rm 8-1000\,\mu m}$ & 80 \\
  $L_{\rm H_2 O}$ & $0.5 \times 10^{-7}$ \\
  \hline
  \hline
 \end{tabular}~\par
$^\dag$Ranges refer to the absorption-corrected values for the AGN continuum between models M and T.
\end{center}
\end{table}

\subsection{On the origin of the soft X-ray emission}
\label{sec:soft}

\begin{figure*}
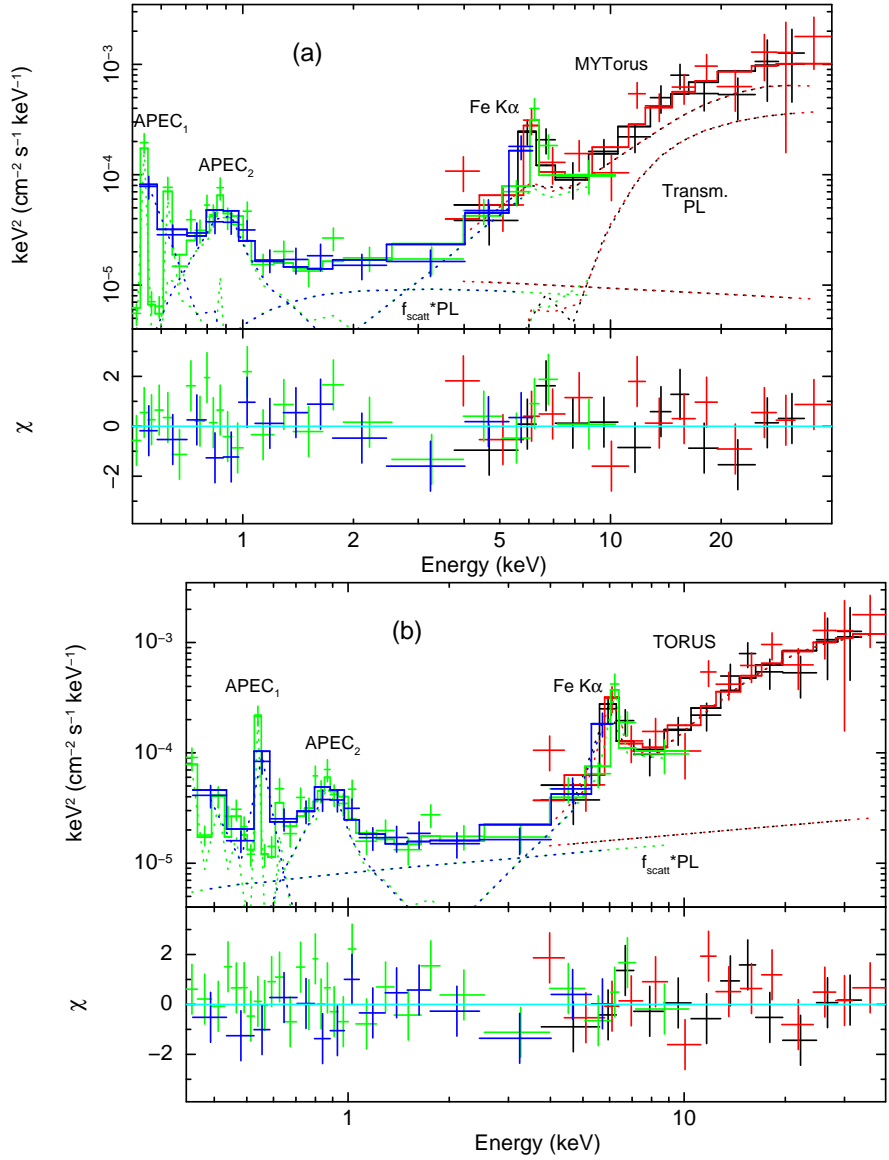

  \begin{center}
  \hspace*{-1.5cm}
  \includegraphics[angle=270,width=11.5cm]{f2a.ps}
  \includegraphics[angle=270,width=11.5cm]{f2b.ps}
  \caption{X-ray fits to the combined \nustar\ and \xmm\ data. Shown in $E F_E$ units are the unfolded model M {\bf (a)} and model T {\bf (b)} fits, respectively. The bottom panels show the residuals in terms of sigmas with error bars of size one. The transmitted PL is not treated separately in the {\sc torus} model T. Color scheme: black (FPMA), red (FPMB), green (pn), blue (MOS1+2). The \mytorus\ fit was restricted to energies above 0.5 keV, which is the lower end of the range over which it is defined. 
    \label{fig:mainfit}}
  \end{center}
\end{figure*}

The soft X-ray emission of Mrk~34 is interesting because of the presence of high-luminosity components which steepen the overall source spectrum significantly below $\sim$3 keV (Fig.~\ref{fig:mainfit}). In fact, the first published X-ray detection of Mrk~34 appears to have been from the \rosat\ High Resolution Imager sensitive over the range 0.1--2.4 keV, rather than at harder X-rays \citep{pfefferkorn01}. 

Two \apec\ components are used to parametrize the soft X-rays below 1 keV, and a PL simulating scattering by diffuse plasma dominates around 2 keV (Fig.~\ref{fig:mainfit}; Table~\ref{tab:xspec}). The presence of multiple hot gas components is common in many systems, especially starbursts \citep[e.g. ][]{konami11, mineo12_hotgas}. However, the power of these components is very large in Mrk~34 and converting the soft band power to a star-formation rate (SFR) using the $L_{0.5-2}$:SFR scaling relation from \citet{mineo12_hotgas} implies a high SFR$_{\rm X-ray} > 140$~\Msunpyr\ even accounting for the scatter in the relation.\footnote{The relation by \citet{mineo12_hotgas} actually assumes a {\sc mekal} model instead of \apec, but there is no significant difference between these two in terms of the inferred luminosities for our data.} One can compare this to the infrared-derived SFR$_{\rm IR}$ from the SFR:$L_{8-1000}$ relation presented by \citet{kennicutt98}. We find SFR$_{\rm IR}\approx~30(\pm 9)$~\Msunpyr, which should be considered as an upper-limit because the AGN contribution to the infrared appears to be substantial. SFR$_{\rm X-ray}$ is much higher than SFR$_{\rm IR}$, suggesting that some other process may be powering the observed soft X-ray components. 

Photoionization of circumnuclear gas is commonly observed in nearby AGN \citep[e.g. ][]{guainazzi07} and is the most likely viable alternative. In the bright nearby galaxy NGC~4151, a photoionized X-ray component with similar fractional luminosity to hard X-rays as in Mrk~34 has been observed to spatially trace the extended \oiii\ emission \citep{wang11_ngc4151}. It may thus be the case that the entire soft X-ray regime in Mrk~34 is a complex of emission lines associated with extended photoionized gas in the narrow line region. We attempted such a model by replacing the \apec\ components with several narrow emission features. A minimal set to give an acceptable fit included \ovii\,K$\alpha$ (0.571 keV), a narrow \oviii\, radiative recombination continuum (0.871 keV), and \nex\,K$\alpha$ (1.022 keV), with $\chi^2$/dof\,=\,68.9/63. The hard X-ray portion of the spectrum is still well fit in this case, but a much steeper scattered PL with $\Gamma$\,=\,2.7\,\p\,0.2 is required to fill in the gaps between the emission lines in the soft band. This indicates the need for extra emission lines, or possibly a weaker collisionally ionized component as has been observed in other AGN \citep{guainazzi09, bianchi10}. 

If photoionization dominates over scattering, this could explain the low $f_{\rm scatt}$$\ltsim$0.5\%\ (Table~\ref{tab:xspec}) associated with the intrinsic PL scattered into the line-of-sight by diffuse hot gas.  Nearby Seyferts generally show $f_{\rm scatt}$ values of a few per cent (e.g. \citealt{cappi06}). If part of this soft emission were instead attributed to X-ray binaries in the host galaxy, that would imply an even lower value of $f_{\rm scatt}$. Alternative explanations could be a \lq buried AGN\rq\ with a geometrically thick torus \citep{ueda07} or host galaxy extinction \citep[][]{hoenig14}, although neither obviously works for Mrk~34. The strong \oiii\ lines seen in this object are not typical of buried AGN, and the small Balmer decrement (see previous section) together with the intermediate inclination angle\footnote{{\tt http://leda.univ-lyon1.fr/}} of the galaxy disk of 45~deg is consistent with only modest dust reddening in the host. Finally, another recent proposal on the origin of this soft emission component is scattering off clouds within a clumpy torus medium \citep{miniutti14}.

These issues can be investigated using higher spatial resolution imaging with \ch, and future high spectral resolution observations with \astroh\ \citep{astroh12}.

\subsection{Comparison to other bona fide CT AGN}
\label{sec:localctagn}

Fig.~\ref{fig:lumdist} compares Mrk~34 on the $L_{\rm 2-10,in}$ vs. distance plane with all other bona fide CT AGN from the compilation of \citet{goulding12}, supplemented with recent results on individual sources which have been collated in the Appendix. Some sources known to be of changing-look nature with rapid Compton-thick/thin transitions such as NGC\,1365 \citep[][]{risaliti05} have not been tabulated in this list in accordance with \citet{dellaceca08}, though these are also potential candidates for inclusion. Note that Mrk~231 and NGC~7674 have been removed from this compilation based upon results from \citet{teng14} and \citet{bianchi05}. Recent \nustar\ observations of the former source show no evidence of CT columns, while the nature of obscuration in the latter is currently unclear with the source being either a changing-look AGN or having recently switched off. For another source, Superantennae, \nustar\ observations find a strong decrease in the broadband X-ray flux as compared to previous observations with \suzaku. Whether this is a result of dramatic luminosity decline, spectral change, or contaminating sources in previous data, is currently being investigated (Teng et al. 2014, in prep.), so this source is also currently not included in this compilation. 

Mrk~34 emerges as one of the most luminous amongst the local bona fide CT AGN, similar to NGC~6240. Given the extreme luminosities of these two sources within the local sample, it is interesting to compare and contrast their characteristics. For NGC~6240, we have reanalyzed the \sax\ data of \citet{vignati99} as well as new \nustar\ data, and found a luminosity agreeing to within a factor of $\sim$2 of the value of $L_{\rm 2-10, in}$ quoted in \citeauthor{vignati99} after correcting for cosmology and applying a correction factor for a toroidal obscuration geometry (Puccetti et al. 2014, in prep.). NGC\,6240 appears to be somewhat less obscured and has a weaker Fe line than Mrk~34 (\nh$\approx$1$\times$10$^{24}$ cm$^{-2}$ and EW$_{\rm K\alpha}$$\approx$0.3 keV; \citealt{brightmannandra11}). 

NGC~6240 is a well-known binary AGN \citep{komossa03} and its hard X-ray emission above 10 keV is the combined luminosity of both the northern (N) and the southern (S) components. The contribution of each component to the hard X-ray emission has not been directly resolved, but low energy X-ray data, as well as mid-IR continuum imaging, show S to be brighter than N by a factors of a few \citep{komossa03, wang14_n6240, asmus14}. Mrk~34, on the other hand, is not a known binary AGN and is not in a major merger. Fig.~\ref{fig:sdssfindingchart} presents \hst\ images for the two objects which makes the difference immediately clear. NGC~6240 has a total far-infrared power $L_{8-1000}$$\approx$6.5$\times$10$^{11}$~\Lsun\ \citep{sanders03_rbgs, koss13}, which is about three times higher than that of Mrk~34. Both objects lie within the luminosity regime associated with LIRGs. We also note that the object that follows these two in terms of luminosity CGCG~420--15 \citep{severgnini11}, has also been identified to lie in a group \citep{crook07}. In this case, we quote the intrinsic luminosity based upon our model fits to new \nustar\ data. 

Is there any evidence for galaxy interaction in Mrk~34? The galaxy is classified morphologically as Hubble class Sa \citep{nair10}. From integral field spectroscopy, \citet{stoklasova09} found an asymmetry in the nuclear emission line distribution which is elongated at a position angle of 140--150~deg. Twisted \ha\ velocity isocontours were also identified by them. However, these may be signatures of interaction between the AGN jet and narrow line region clouds, as suggested by \citet{falcke98} based upon matching of radio and narrowband emission line maps. Any morphological perturbations related to galaxy interactions must then be relatively mild. Even if the present nuclear activity is a result of some past interaction or merger, Mrk~34 is clearly at a different evolutionary stage now as compared to the strongly interacting system NGC~6240. Alternatively, these sources may be showcasing the contrast between the two dominant AGN-triggering modes -- secular processes vs. major mergers -- with secular processes thought to dominate in the low-redshift universe \citep[e.g. ][]{draperballantyne12_secular}. 

In summary, Mrk~34 is not in a merging system at present, unlike NGC~6240. This makes Mrk~34 the closest known bona fide CT QSO in a non-merging system.

\subsection{Feedback from the AGN?}

At present, there is an absence of nuclear star formation in Mrk~34 \citep{gonzalezdelgado01,stoklasova09}. \citet{wang07_feedback} classify Mrk~34 as undergoing suppressed star formation as a result of AGN feedback, with the present SFR surface density lying about two orders or magnitude lower than predicted by the Kennicutt-Schmidt Law \citep{kennicutt98_ks}. 

A spatially resolved ionized outflow with a maximum velocity of 1500~km~s$^{-1}$ and gas being accelerated out to a radius of 1 kpc has been observed in high angular resolution \hst\ observations with the Space Telescope Imaging Spectrograph by \citet{fischer13}. It is possible to estimate the potential impact of mechanical feedback associated with this ionized \oiii\ outflow. For this, we use the relation between \loiii\ (assuming that the line is fully associated with the outflow) and the kinetic power ($P_{\rm K}^{\rm ion}$) of this ionized component, as derived by \citet{canodiaz12}. For typical electron densities of 10$^{2-4}$\,cm$^{-3}$ expected in the narrow line region, Eq.~B.9 of \citet{canodiaz12} implies a kinetic power of $P_{\rm K}^{\rm ion} \sim 4\times 10^{41-43}$\,erg\,s$^{-1}$. This constitutes $\sim 0.05-5$\,\% of the source (infrared) bolometric power, a typical level required by models for the outflow to have any significant impact on the host galaxy \citep[e.g. ][]{dimatteo05, hopkinselvis10}. 

As an alternative determination of the outflow power, we utilize the \hb\ luminosity together with Eqs. 2 and 3 of \citet{harrison14}. The line luminosity ($L_{\rm H\beta}$\,=\,$2\times 10^{41}$\,erg\,s$^{-1}$) and width containing 80\%\ of the flux ($W_{80}$\,=690\,km\,s$^{-1}$) were measured from the SDSS spectra as analysed in \citet{mullaney13}. In this case, the SDSS observations do not spatially resolve the line emission so we assume the extent and maximum speed of the \hb\ line emitting material to be the same as that of the ionized gas \citep{fischer13}. These assumptions imply that the outflow has been constant for $\approx$\,1\,Myr, which results in an ionized gas mass $M_{\rm gas}\sim 6\times 10^{5-7}$\,\Msun, a kinetic energy $E_{\rm kin}\sim 2\times 10^{54-56}$\,erg, and kinetic power $\dot{E}_{\rm kin}\sim 4\times 10^{40-42}$\,erg\,s$^{-1}$, a range that is about one order of magnitude lower than the determination based upon \citet{canodiaz12}.

The above estimates use observed line luminosities, and a correction for reddening would push the corresponding power higher by factors of a few. On the other hand, we assume that the line radiating ionized gas fully participates in the outflow, which is unlikely to be the case. For example, in the outflow estimates based upon \hb, we incorporate the total observed narrow line flux rather than the clearly outflowing (blueshifted) components. These estimates would then provide upper limits on the outflow, though spherical and wide-angle outflows with small velocity differences across them have been observed in many systems \citep[e.g. ][]{liu13, harrison14}. So mechanical feedback may have a significant effect in Mrk~34, though there are plenty of uncertainties in the present estimates. 

In any case, the presence of CT gas columns as inferred from the \nustar\ data means that such feedback has not yet removed the nuclear obscuring gas. Any jet-induced interaction is expected to have most impact along the jet axis. The source has a bipolar radio morphology and shows two hot spots \citep{falcke98}, so the jet is pointed out of our line-of-sight and would not directly impact the toroidal obscuring gas. Direct radiation pressure is also unlikely to be effective in removing obscuring gas when the column density become Compton-thick \citep[e.g. ][]{fabian08}. But radiation pressure could be the physical driver of the ionized outflow.

\subsection{Implications for distant obscured AGN studies} 

\begin{figure*}
  \begin{center}
  \includegraphics[width=11.5cm,angle=90]{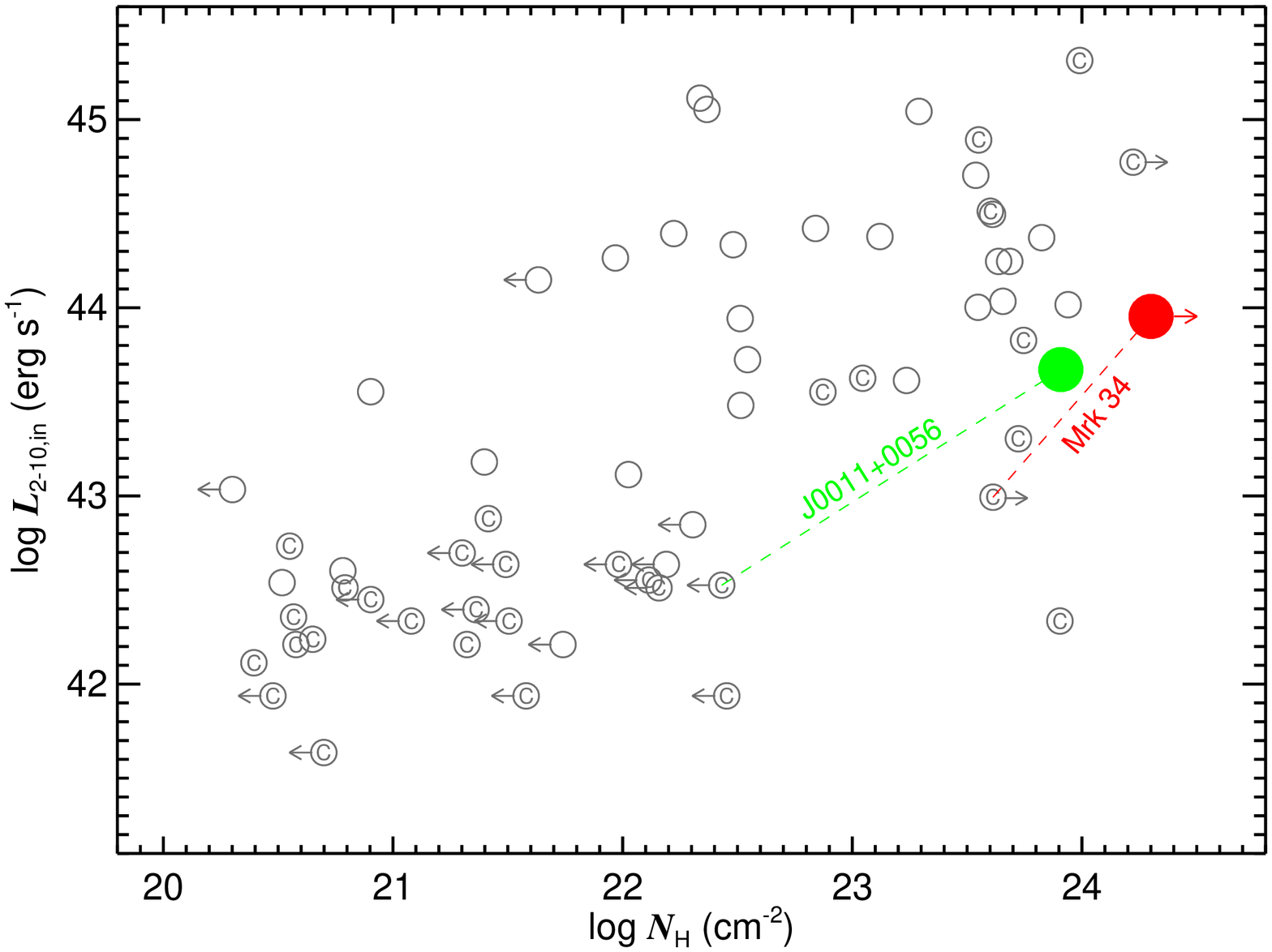}
  \caption{\nh\ vs. $L_{\rm 2-10,in}$ for all X-ray--detected SDSS QSO2s (gray), based only upon direct continuum modeling of X-ray data below 10 keV, collated from \citet{vignali06, vignali10}, \citet{jia13} and \citet{lamassa14}. \nh\ upper limits are denoted by arrows and CT candidates are denoted with \lq C\rq. Recall that CT candidacy is often based upon indirect multiwavelength indicators, and direct continuum fitting to low-energy X-ray data of faint, obscured AGN is very inefficient at securely identifying CT sources. This is why many CT candidates have apparent \nh$<<$10$^{24}$ cm$^{-2}$ at present, and their plotted values of $L_{\rm 2-10,in}$ are also likely to be underestimates. The large filled points are the updated values from \nustar\ modeling for the two \nustar-detected sources (Mrk~34 in red and SDSS~J0011+0056 in green, respectively, where the latter is reported in \citealt{lansbury14}). 
    \label{fig:nhlum}}
  \end{center}
\end{figure*}

Fig.~\ref{fig:nhlum} shows our present constraints on \nh\ and the intrinsic \lx\ for the X-ray--detected SDSS-QSO2 population. In gray are the measurements for all 63 sources detected by \ch/\xmm\ and with an \nh\ measurement based upon X-ray spectral fitting with data below 10 keV, collated from \citet{vignali06, vignali10}, \citet{jia13} and \citet{lamassa14}. For sources with multiple fits (either to data from multiple observations, or from using multiple models), we chose fits based upon recent physically-motivated torus models when available, otherwise preferring the highest fitted \nh\ values. In red and green are the new constraints from the broadband modeling of the two \nustar--detected QSO2s Mrk~34 and SDSS~J0011+0056, respectively. Direct continuum modeling of data below 10 keV underestimates \nh\ and intrinsic \lx\ for both, and the improved constraints enabled by \nustar\ move both quantities to significantly higher values. So, the plotted values of \nh\ and $L_{\rm 2-10,in}$ may actually be {\em underestimates} for other sources as well, especially the CT candidates. If this trend turns out to hold for many of these candidates, this will have important implications for AGN population models which require a good knowledge of the underlying column density and luminosity distributions. Observations of more QSO2s with \nustar\ would be invaluable for obtaining improved constraints on the overall population. 

Studies of distant obscured AGN are extremely challenging even with \nustar, as is evident from the non-detection of all sources except one in two recent exploratory \nustar\ studies targeting luminous QSO2s at $z \sim 0.5$ \citep{lansbury14} and hyperluminous infrared galaxies at $z\sim2$ \citep{stern14}, respectively. Mrk~34 is the first source in a similar class to show enough photons for X-ray spectral modeling. Its brightness is a result of its lower redshift as compared to the previously targeted sources. This, together with the fact that Mrk~34 appears to be a typical CT QSO (as discussed before), means that the X-ray spectrum of Mrk\,34 can serve as a useful template for more distant CT QSOs. In particular, Mrk\,34 is a factor of 2--3 more luminous than NGC\,1068, and shows evidence of a higher level of obscuration as compared to NGC\,6240 (see section~\ref{sec:localctagn}), the two other sources often considered as CT AGN archetypes. 

An important point is that Mrk~34 would not have been {\em selected} as an obscured quasar from modeling of the X-ray continuum below 10 keV alone, because absorption-correction of the \xmm\ data without higher energy coverage results in intrinsic luminosities that are $\gtsim$10 times lower than found by \nustar\ (\S~\ref{sec:sampleselection}). This means that X-ray selection of obscured quasars from data with low spectral statistics can severely underestimate \nh\ as well as the intrinsic X-ray power, leading to biased estimates of the distributions of these quantities. While the high EW$_{\rm K\alpha}$ in Mrk~34 was known previously and gave a strong hint supporting CT obscuration, detection and identification of this narrow feature requires a reasonable count-rate, which is not available for most AGN found in typical distant X-ray surveys. Expanded deep surveys with \ch\ and \xmm\ (e.g. the upcoming 7~Ms Chandra Deep Field South) will help in this regard, as will detailed follow-up with future missions such as \astroh\ and \athena\ \citep{astroh12, athenaplus}.

\section{Summary}

Combining \nustar\ observations with archival \xmm\ data, we have carried out high quality broadband X-ray spectroscopy of the optically selected Type 2 quasar Mrk~34. A summary of our main results is as follows: 

\begin{enumerate}

\item Using two physically motivated toroidal obscuration models, we show that the spectra are fully consistent with obscuration by Compton-thick column densities of gas along the line-of-sight. This is the first time that such high columns have been directly measured from spectral fitting to the X-ray continuum of an SDSS-QSO2. This has been possible thanks to the hard X-ray sensitivity of \nustar. 

\item Comparisons of the intrinsic luminosity inferred from the two torus models with various multiwavelength luminosity correlations suggests that our X-ray analysis reliably measures the intrinsic source power of $L_{\rm 2-10,\ in}\sim 10^{44}$~erg~s$^{-1}$ to within a factor of $\sim 2$. When converted to a bolometric luminosity, the accretion power is sufficient to drive the entire infrared emission. 

\item The observed soft X-ray emission appears to be too luminous to be associated with star-formation and may instead be driven by AGN photoionization. 

\item Mrk~34 is representative of the SDSS-QSO2 population in terms of its \oiii\ line luminosity, and it shows all the indirect pieces of evidence expected for sources with Compton-thick obscuration (i.e., a low observed X-ray luminosity with respect to other isotropic AGN luminosity indicators, a powerful H$_2$O megamaser, and a strong Fe K$\alpha$ line). Thus, Mrk~34 is a benchmark CT QSO2 in the local universe. It is the nearest isolated CT quasar, in that it is not presently undergoing a major merger. 

\item Yet, using X-ray data below 10 keV alone fails to pick up the source as an intrinsically luminous AGN. This has implications for low-energy X-ray selection of obscured quasars in survey fields where most detected sources lie in the low-count regime, and more \nustar\ observations of CT QSO2 candidates are required for understanding the importance of this selection effect. Our broadband X-ray spectrum of Mrk~34 could serve as a useful local template for hard X-ray studies of distant CT quasars. 

\end{enumerate}

\begin{figure*}
  \begin{center}
  \includegraphics[angle=90,width=16cm]{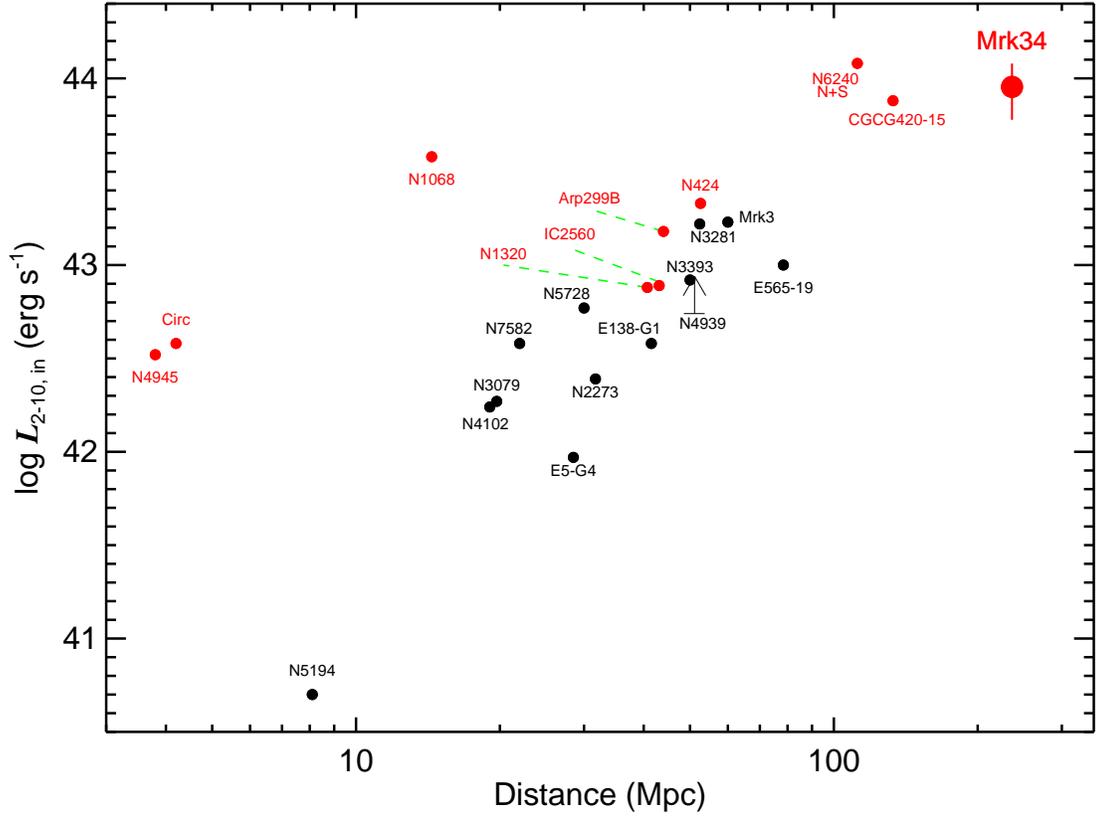}
  \caption{Intrinsic 2--10 keV luminosity vs. distance for the sample of bona fide CT AGN from \citet{goulding12}, updated and supplemented with other recent results. The data used to make this plot are collated in Table~\ref{tab:ctagn}. Red denotes constraints from \nustar. 
    \label{fig:lumdist}}
  \end{center}
\end{figure*}

\begin{figure*}
  \begin{center}
  \includegraphics[width=16cm]{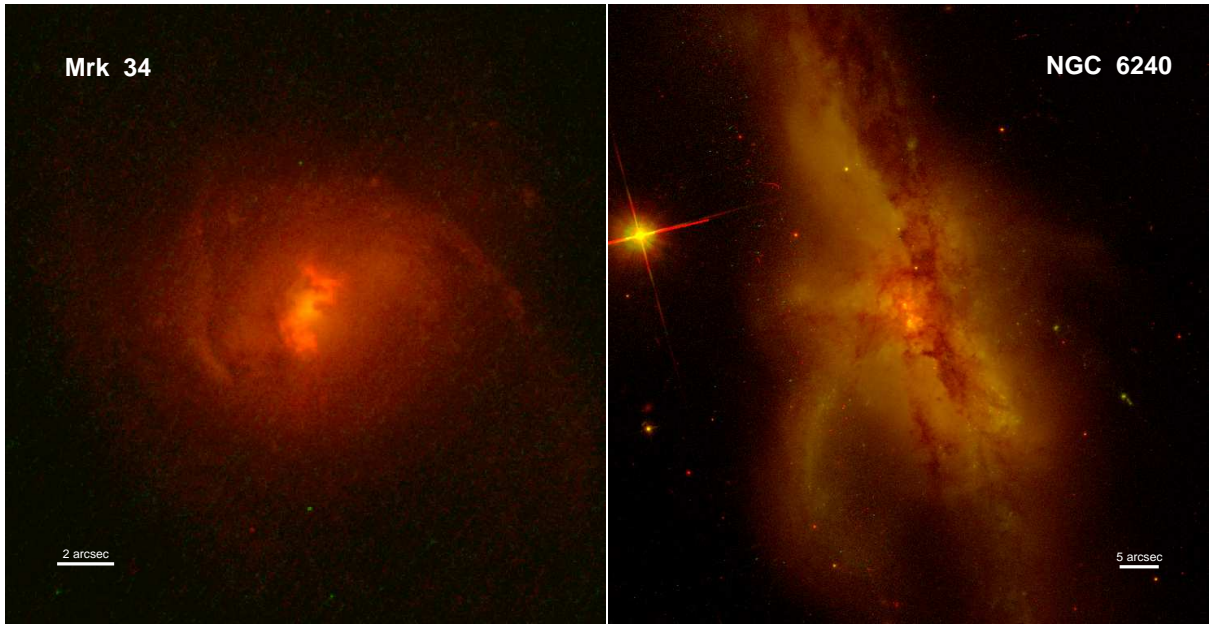}
  \caption{Hubble Legacy Archive images of Mrk~34 and NGC~6240. The images are two-band color composites. For Mrk~34, WFPC2 images in filters F547M (red) and F467M (green) are combined; for NGC~6240, ACS filters used are F814W (red) and F435W (green). North is up and East to the left in both panels. These images clearly show the relatively unperturbed disk morphology of Mrk~34 as compared to the strong ongoing interaction in NGC~6240. 
    \label{fig:sdssfindingchart}}
  \end{center}
\end{figure*}

\acknowledgements
Grant and fellowship acknowledgments: STFC ST/J003697/1 (P.G.), ST/K501979/1 (G.B.L.), ST/I001573/1 (D.M.A. and A.D.M.), Leverhulme Trust (D.M.A.), NASA Postdoctoral Program (S.H.T), ASI-INAF grant (A.C.), Anillo ACT1101 and FONDECYT 1140304 (P.A.), International Fulbright Science and Technology Award (M.B.), and Swiss National Science Foundation (NSF) grant PP00P2 138979/1 (M.K.). In addition, F.A.H. acknowledges support from a Durham University COFUND fellowship, and F.E.B. acknowledges support from Basal-CATA PFB-06/2007, CONICYT-Chile (FONDECYT 1141218 and "EMBIGGEN" Anillo ACT1101) Project IC120009 "Millennium Institute of Astrophysics (MAS)" funded by the Iniciativa Cient\'{\i}fica Milenio del Ministerio de Econom\'{\i}a, Fomento y Turismo. The authors thank Fred K.Y. Lo for megamaser discussions, and the referee for their report. P.G. thanks James R. Mullaney and Chris M. Harrison for discussions. P.G. is also grateful to Matteo Guainazzi for his comments and insights on the origin of the soft X-ray spectrum. 

\nustar\ is a project led by the California Institute of Technology (Caltech), managed by the Jet Propulsion Laboratory (JPL), and funded by the National Aeronautics and Space Administration (NASA). The \nustar\ Operations, Software and Calibration teams are acknowledged for support with these observations. This research has made use of the \nustar\ Data Analysis Software ({\sc nustardas}) jointly developed by the ASI Science Data Center (ASDC, Italy) and the California Institute of Technology (USA). This work has made use of data from \xmm\ and the Sloan Digital Sky Survey. Fig.~\ref{fig:sdssfindingchart} is based upon data from with the NASA/ESA Hubble Space Telescope, and obtained from the Hubble Legacy Archive. This research has made use of the NASA/IPAC Extragalactic Database (NED), which is operated by JPL, caltech, under contract with NASA.

\facility{{\em Facilities:} \nustar, \xmm, Sloan, {\em WISE}, {\em HST}, \swift, \ch.}

\begin{table*}[h]
  \begin{center}
  \caption{X-ray spectral fits\label{tab:xspec}}
  \begin{tabular}{lcccr}
\hline
Component          &  Parameter       &     Model M     &    Model T   & Units \\
\hline
\hline
\apecone           & $kT_1$           & $0.18 \pm 0.03$ & $0.11_{-0.02}^{+0.03}$ & keV\\
                   & $L_{0.5-2}$       & 2.0 & 1.6 &  $\times$ 10$^{41}$ erg s$^{-1}$\\
\apectwo           & $kT_2$           & $0.96 \pm 0.10$ & $0.93_{-0.13}^{+0.07}$ &  keV\\
                   & $L_{0.5-2}$        & 2.4 & 2.2 & $\times$ 10$^{41}$ erg s$^{-1}$\\
\mytorus/\torus    & \nh(eq)          & $9.51_{-4.17}^{+u}$ &   --              &  $\times$ 10$^{24}$ cm$^{-2}$\\
                   & \nh(los)         & $2.45_{-1.08}^{+u}$ & 35.5$_{-31.9}^{+u}$  &  $\times$ 10$^{24}$ cm$^{-2}$\\
                   & $\theta_{\rm inc}$         & $61_{-0.5}^{+14}$    & $87^f$ & deg\\
                   & $\theta_{\rm tor}$         & --               & $67_{-40}^{+11}$ & deg\\
                   & EW(Fe K$\alpha$) & 1.2 & 1.4 &  keV \\
AGN continuum      & $\Gamma$         & $2.2_{-0.3}^{+0.2}$ & $1.7_{-0.5}^{+0.4}$ &  \\
                   & $L_{2-10}$         & 0.6 & 1.2 &  $\times$ 10$^{44}$ erg s$^{-1}$\\
                   & $L_{0.5-30}$       & 1.5 & 3.1  &  $\times$ 10$^{44}$ erg s$^{-1}$\\
Diffuse Scattering & \fscatt          & $3.5_{-1.4}^{+2.2}$& $2.2_{-1.8}^{+2.1}$ &  $\times$ 10$^{-3}$\\
Large-scale absorption & \nh(host)    & $4.4_{-0.1}^{+0.4}$& -- & $\times$ 10$^{21}$ cm$^{-2}$\\
{\em XMM}:\nustar\ cross-calib & {\sc const}   & $0.84_{-0.20}^{+0.32}$ & $0.83_{-0.16}^{+0.21}$ & \\
                   &                  & & &  \\
$\chi^2$/dof       &                  & 57.5/53 & 59.9/62 & \\
\hline
\end{tabular}~\par
$^u$unconstrained. $^f$fixed.\\
Model M: \mytorus\ coupled component \citep{mytorus} used for the circumnuclear absorber/reflector.\\
Model T: \torus\ component \citep{brightmannandra11} used for the circumnuclear absorber/reflector.
\end{center}
\end{table*}

\clearpage
\appendix
\section{List of {\em bona fide} local Compton-thick AGN}

Table~\ref{tab:ctagn} lists the distances and intrinsic luminosities of all the bona fide Compton-thick AGN plotted in Fig.~\ref{fig:lumdist}, along with relevant references for their X-ray analyses. 

\begin{table}
  \begin{center}
  \caption{List of {\em bona fide} local Compton-thick AGN\label{tab:ctagn}}
  \begin{tabular}{lccr}
\hline
Source          & Distance   & $L_{\rm 2-10,in}$ & Reference \\
                &  Mpc       &    erg s$^{-1}$ & \\ 
\hline
\hline
        NGC 424 &  52.6 &    43.33 &   1\\
       NGC 1068 &  14.4 &    43.58 &   2\\
       NGC 1320 &  40.7 &    42.88 &   1\\
    CGCG420--15 & 133.0 &    43.88 &   3, 4\\
  ESO 005--G004 &  28.5 &    41.97 &   \\
          Mrk 3 &  60.0 &    43.23 &   5\\
       NGC 2273 &  31.7 &    42.39 &   \\
  ESO 565--G019 &  78.4 &    43.00 &   6\\
       NGC 3079 &  19.7 &    42.27 &   \\
        IC 2560 &  43.1 &    42.89 &   1\\
       NGC 3281 &  52.4 &    43.22 &   \\
         Mrk 34 & 236.0 &    43.95 &   4\\
       NGC 3393 &  50.0 &    42.92 &   7\\
       Arp 299B &  44.0 &    43.18 &   8\\
       NGC 4102 &  19.0 &    42.24 &   9\\
       NGC 4939 &  51.1 &    42.74 &   \\
       NGC 4945 &   3.8 &    42.52 &   10\\
       NGC 5194 &   8.1 &    40.70 &   \\
       Circinus &   4.2 &    42.58 &   11\\
       NGC 5728 &  30.0 &    42.77 &   \\
  ESO 138--G001 &  41.5 &    42.58 &   \\
       NGC 6240 & 112.0 &    44.08 &   12, 13\\
       NGC 7582 &  22.0 &    42.58 &   \\
\hline
\end{tabular}~\par
Distances are redshift-independent estimates from NED for the closest sources, or luminosity distances from the respective references, corrected for cosmology. 
{\bf References.} (1) \citet{balokovic14}; (2) Bauer et al. (2014, in prep); (3) \citet{severgnini11}; (4) this work; (5) \citet{awaki08}; (6) \citet{g13_eso565}; (7) \citet{fabbiano11}; (8) \citet{ptak14}; (9) \citet{gonzalezmartin11}; (10) \citet{puccetti14}; (11) \citet{arevalo14}; (12) \citet{vignati99}; (13) Puccetti et al. (2014, in prep.). 
Where not stated, the reference is the compilation by \cite{goulding12} and papers referred to therein. Mrk~231, NGC~7674 and IRAS~19254--72 are not included as a result of recent updates to the intrinsic luminosities (see text). 
\end{center}
\end{table}

\label{lastpage}

\begin{thebibliography}{137}
\expandafter\ifx\csname natexlab\endcsname\relax\def\natexlab#1{#1}\fi

\bibitem[{{Akylas} {et~al.}(2012){Akylas}, {Georgakakis}, {Georgantopoulos},
  {Brightman}, \& {Nandra}}]{akylas12}
{Akylas}, A., {Georgakakis}, A., {Georgantopoulos}, I., {Brightman}, M., \&
  {Nandra}, K. 2012, \aap, 546, A98, arXiv:1209.5398

\bibitem[{{Alexander} {et~al.}(2013){Alexander}, {Stern}, {Del Moro},
  {Lansbury}, {Assef}, {Aird}, {Ajello}, {Ballantyne}, {Bauer}, {Boggs},
  {Brandt}, {Christensen}, {Civano}, {Comastri}, {Craig}, {Elvis},
  {Grefenstette}, {Hailey}, {Harrison}, {Hickox}, {Luo}, {Madsen}, {Mullaney},
  {Perri}, {Puccetti}, {Saez}, {Treister}, {Urry}, {Zhang}, {Bridge},
  {Eisenhardt}, {Gonzalez}, {Miller}, \& {Tsai}}]{alexander13}
{Alexander}, D.~M. {et~al.} 2013, \apj, 773, 125, arXiv:1307.1733

\bibitem[{{Alonso-Herrero} {et~al.}(2012){Alonso-Herrero}, {Pereira-Santaella},
  {Rieke}, \& {Rigopoulou}}]{alonsoherrero12}
{Alonso-Herrero}, A., {Pereira-Santaella}, M., {Rieke}, G.~H., \& {Rigopoulou},
  D. 2012, \apj, 744, 2, 1109.1372

\bibitem[{{Ar{\'e}valo} {et~al.}(2014){Ar{\'e}valo}, {Bauer}, {Puccetti},
  {Walton}, {Koss}, {Boggs}, {Brandt}, {Brightman}, {Christensen}, {Comastri},
  {Craig}, {Fuerst}, {Gandhi}, {Grefenstette}, {Hailey}, {Harrison}, {Luo},
  {Madejski}, {Madsen}, {Marinucci}, {Matt}, {Saez}, {Stern}, {Stuhlinger},
  {Treister}, {Urry}, \& {Zhang}}]{arevalo14}
{Ar{\'e}valo}, P. {et~al.} 2014, ApJ in press, arXiv:1406.3345

\bibitem[{{Arnaud}(1996)}]{xspec}
{Arnaud}, K.~A. 1996, in ASP Conf. Ser. 101: Astronomical Data Analysis
  Software and Systems V, eds. George H. Jacoby and Jeannette Barnes, Vol.~5,
  17--+

\bibitem[{{Asmus} {et~al.}(2011){Asmus}, {Gandhi}, {Smette}, {H{\"o}nig}, \&
  {Duschl}}]{asmus11}
{Asmus}, D., {Gandhi}, P., {Smette}, A., {H{\"o}nig}, S.~F., \& {Duschl}, W.~J.
  2011, \aap, 536, A36, 1109.4873

\bibitem[{{Asmus} {et~al.}(2014){Asmus}, {H{\"o}nig}, {Gandhi}, {Smette}, \&
  {Duschl}}]{asmus14}
{Asmus}, D., {H{\"o}nig}, S.~F., {Gandhi}, P., {Smette}, A., \& {Duschl}, W.~J.
  2014, \mnras, 439, 1648, 1310.2770

\bibitem[{{Assef} {et~al.}(2013){Assef}, {Stern}, {Kochanek}, {Blain},
  {Brodwin}, {Brown}, {Donoso}, {Eisenhardt}, {Jannuzi}, {Jarrett}, {Stanford},
  {Tsai}, {Wu}, \& {Yan}}]{assef13}
{Assef}, R.~J. {et~al.} 2013, \apj, 772, 26

\bibitem[{{Awaki} {et~al.}(2008){Awaki}, {Anabuki}, {Fukazawa}, {Gallo},
  {Ikeda}, {Isobe}, {Itoh}, {Kunieda}, {Makishima}, {Markowitz}, {Miniutti},
  {Mizuno}, {Okajima}, {Ptak}, {Reeves}, {Takahashi}, {Terashima}, \&
  {Yaqoob}}]{awaki08}
{Awaki}, H. {et~al.} 2008, \pasj, 60, 293, arXiv:0707.2425

\bibitem[{{Ballantyne} {et~al.}(2011){Ballantyne}, {Draper}, {Madsen}, {Rigby},
  \& {Treister}}]{ballantyne11}
{Ballantyne}, D.~R., {Draper}, A.~R., {Madsen}, K.~K., {Rigby}, J.~R., \&
  {Treister}, E. 2011, \apj, 736, 56, arXiv:1105.0965

\bibitem[{{Balokovi\'{c} et al.}(2014)}]{balokovic14}
{Balokovi\'{c} et al.} 2014, ApJ submitted

\bibitem[{{Bassani} {et~al.}(1999){Bassani}, {Dadina}, {Maiolino}, {Salvati},
  {Risaliti}, {della Ceca}, {Matt}, \& {Zamorani}}]{bassani99}
{Bassani}, L., {Dadina}, M., {Maiolino}, R., {Salvati}, M., {Risaliti}, G.,
  {della Ceca}, R., {Matt}, G., \& {Zamorani}, G. 1999, \apjs, 121, 473

\bibitem[{{Baumgartner} {et~al.}(2013){Baumgartner}, {Tueller}, {Markwardt},
  {Skinner}, {Barthelmy}, {Mushotzky}, {Evans}, \& {Gehrels}}]{baumgartner13}
{Baumgartner}, W.~H., {Tueller}, J., {Markwardt}, C.~B., {Skinner}, G.~K.,
  {Barthelmy}, S., {Mushotzky}, R.~F., {Evans}, P.~A., \& {Gehrels}, N. 2013,
  \apjs, 207, 19, arXiv:1212.3336

\bibitem[{{Bianchi} {et~al.}(2010){Bianchi}, {Chiaberge}, {Evans}, {Guainazzi},
  {Baldi}, {Matt}, \& {Piconcelli}}]{bianchi10}
{Bianchi}, S., {Chiaberge}, M., {Evans}, D.~A., {Guainazzi}, M., {Baldi},
  R.~D., {Matt}, G., \& {Piconcelli}, E. 2010, \mnras, 405, 553,
  arXiv:1002.0800

\bibitem[{{Bianchi} {et~al.}(2005){Bianchi}, {Guainazzi}, {Matt}, {Chiaberge},
  {Iwasawa}, {Fiore}, \& {Maiolino}}]{bianchi05}
{Bianchi}, S., {Guainazzi}, M., {Matt}, G., {Chiaberge}, M., {Iwasawa}, K.,
  {Fiore}, F., \& {Maiolino}, R. 2005, \aap, 442, 185, astro-ph/0507323

\bibitem[{{Bolton} {et~al.}(2012){Bolton}, {Schlegel}, {Aubourg}, {Bailey},
  {Bhardwaj}, {Brownstein}, {Burles}, {Chen}, {Dawson}, {Eisenstein}, {Gunn},
  {Knapp}, {Loomis}, {Lupton}, {Maraston}, {Muna}, {Myers}, {Olmstead},
  {Padmanabhan}, {P{\^a}ris}, {Percival}, {Petitjean}, {Rockosi}, {Ross},
  {Schneider}, {Shu}, {Strauss}, {Thomas}, {Tremonti}, {Wake}, {Weaver}, \&
  {Wood-Vasey}}]{bolton12}
{Bolton}, A.~S. {et~al.} 2012, \aj, 144, 144, 1207.7326

\bibitem[{{Boroson} \& {Green}(1992)}]{borosongreen92}
{Boroson}, T.~A., \& {Green}, R.~F. 1992, \apjs, 80, 109

\bibitem[{{Brightman} \& {Nandra}(2011)}]{brightmannandra11}
{Brightman}, M., \& {Nandra}, K. 2011, \mnras, 413, 1206, arXiv:1012.3345

\bibitem[{{Brightman} \& {Ueda}(2012)}]{brightmanueda12}
{Brightman}, M., \& {Ueda}, Y. 2012, \mnras, 423, 702, arXiv:1203.1045

\bibitem[{{Burlon} {et~al.}(2011){Burlon}, {Ajello}, {Greiner}, {Comastri},
  {Merloni}, \& {Gehrels}}]{burlon11}
{Burlon}, D., {Ajello}, M., {Greiner}, J., {Comastri}, A., {Merloni}, A., \&
  {Gehrels}, N. 2011, \apj, 728, 58, arXiv:1012.0302

\bibitem[{{Cano-D{\'{\i}}az} {et~al.}(2012){Cano-D{\'{\i}}az}, {Maiolino},
  {Marconi}, {Netzer}, {Shemmer}, \& {Cresci}}]{canodiaz12}
{Cano-D{\'{\i}}az}, M., {Maiolino}, R., {Marconi}, A., {Netzer}, H., {Shemmer},
  O., \& {Cresci}, G. 2012, \aap, 537, L8, 1112.3071

\bibitem[{{Cappi} {et~al.}(2006){Cappi}, {Panessa}, {Bassani}, {Dadina},
  {Dicocco}, {Comastri}, {della Ceca}, {Filippenko}, {Gianotti}, {Ho},
  {Malaguti}, {Mulchaey}, {Palumbo}, {Piconcelli}, {Sargent}, {Stephen},
  {Trifoglio}, \& {Weaver}}]{cappi06}
{Cappi}, M. {et~al.} 2006, \aap, 446, 459, arXiv:astro-ph/0509584

\bibitem[{{Comastri} {et~al.}(2011){Comastri}, {Ranalli}, {Iwasawa}, {Vignali},
  {Gilli}, {Georgantopoulos}, {Barcons}, {Brandt}, {Brunner}, {Brusa},
  {Cappelluti}, {Carrera}, {Civano}, {Fiore}, {Hasinger}, {Mainieri},
  {Merloni}, {Nicastro}, {Paolillo}, {Puccetti}, {Rosati}, {Silverman},
  {Tozzi}, {Zamorani}, {Balestra}, {Bauer}, {Luo}, \& {Xue}}]{comastri11}
{Comastri}, A. {et~al.} 2011, \aap, 526, L9, arXiv:1012.4011

\bibitem[{{Comastri} {et~al.}(1995){Comastri}, {Setti}, {Zamorani}, \&
  {Hasinger}}]{comastri95}
{Comastri}, A., {Setti}, G., {Zamorani}, G., \& {Hasinger}, G. 1995, \aap, 296,
  1+

\bibitem[{{Crook} {et~al.}(2007){Crook}, {Huchra}, {Martimbeau}, {Masters},
  {Jarrett}, \& {Macri}}]{crook07}
{Crook}, A.~C., {Huchra}, J.~P., {Martimbeau}, N., {Masters}, K.~L., {Jarrett},
  T., \& {Macri}, L.~M. 2007, \apj, 655, 790, astro-ph/0610732

\bibitem[{{Dahari} \& {De Robertis}(1988)}]{dahariderobertis88}
{Dahari}, O., \& {De Robertis}, M.~M. 1988, \apjs, 67, 249

\bibitem[{{Del Moro} {et~al.}(2014){Del Moro}, {Mullaney}, {Alexander},
  {Comastri}, {Bauer}, {Treister}, {Stern}, {Civano}, {Ranalli}, {Vignali},
  {Aird}, {Ballantyne}, {Balokovi{\'c}}, {Boggs}, {Brandt}, {Christensen},
  {Craig}, {Gandhi}, {Gilli}, {Hailey}, {Harrison}, {Hickox}, {LaMassa},
  {Lansbury}, {Luo}, {Puccetti}, {Urry}, \& {Zhang}}]{delmoro14}
{Del Moro}, A. {et~al.} 2014, \apj, 786, 16, arXiv:1403.2491

\bibitem[{{Della Ceca} {et~al.}(2008){Della Ceca}, {Severgnini}, {Caccianiga},
  {Comastri}, {Gilli}, {Fiore}, {Piconcelli}, {Malaguti}, \&
  {Vignali}}]{dellaceca08}
{Della Ceca}, R. {et~al.} 2008, \memsai, 79, 65, arXiv:0709.3060

\bibitem[{{Di Matteo} {et~al.}(2005){Di Matteo}, {Springel}, \&
  {Hernquist}}]{dimatteo05}
{Di Matteo}, T., {Springel}, V., \& {Hernquist}, L. 2005, \nat, 433, 604,
  arXiv:astro-ph/0502199

\bibitem[{{Dickey} \& {Lockman}(1990)}]{dickeylongman90}
{Dickey}, J.~M., \& {Lockman}, F.~J. 1990, \araa, 28, 215

\bibitem[{{Done} {et~al.}(2003){Done}, {Madejski}, {{\.Z}ycki}, \&
  {Greenhill}}]{done03}
{Done}, C., {Madejski}, G.~M., {{\.Z}ycki}, P.~T., \& {Greenhill}, L.~J. 2003,
  \apj, 588, 763, arXiv:astro-ph/0301383

\bibitem[{{Donley} {et~al.}(2012){Donley}, {Koekemoer}, {Brusa}, {Capak},
  {Cardamone}, {Civano}, {Ilbert}, {Impey}, {Kartaltepe}, {Miyaji}, {Salvato},
  {Sanders}, {Trump}, \& {Zamorani}}]{donley12}
{Donley}, J.~L. {et~al.} 2012, \apj, 748, 142, 1201.3899

\bibitem[{{Draper} \& {Ballantyne}(2010)}]{draperballantyne10}
{Draper}, A.~R., \& {Ballantyne}, D.~R. 2010, \apjl, 715, L99, 1004.0690

\bibitem[{{Draper} \& {Ballantyne}(2012)}]{draperballantyne12_secular}
------. 2012, \apj, 751, 72, 1203.5117

\bibitem[{{Elvis} {et~al.}(1994){Elvis}, {Wilkes}, {McDowell}, {Green},
  {Bechtold}, {Willner}, {Oey}, {Polomski}, \& {Cutri}}]{elvis94}
{Elvis}, M. {et~al.} 1994, \apjs, 95, 1

\bibitem[{{Fabbiano} {et~al.}(2011){Fabbiano}, {Wang}, {Elvis}, \&
  {Risaliti}}]{fabbiano11}
{Fabbiano}, G., {Wang}, J., {Elvis}, M., \& {Risaliti}, G. 2011, \nat, 477,
  431, arXiv:1109.0483

\bibitem[{{Fabian} {et~al.}(2008){Fabian}, {Vasudevan}, \& {Gandhi}}]{fabian08}
{Fabian}, A.~C., {Vasudevan}, R.~V., \& {Gandhi}, P. 2008, \mnras, 385, L43,
  arXiv:0712.0277

\bibitem[{{Falcke} {et~al.}(1998){Falcke}, {Wilson}, \& {Simpson}}]{falcke98}
{Falcke}, H., {Wilson}, A.~S., \& {Simpson}, C. 1998, \apj, 502, 199,
  astro-ph/9801086

\bibitem[{{Feruglio} {et~al.}(2011){Feruglio}, {Daddi}, {Fiore}, {Alexander},
  {Piconcelli}, \& {Malacaria}}]{feruglio11}
{Feruglio}, C., {Daddi}, E., {Fiore}, F., {Alexander}, D.~M., {Piconcelli}, E.,
  \& {Malacaria}, C. 2011, \apjl, 729, L4, 1101.3478

\bibitem[{{Fischer} {et~al.}(2013){Fischer}, {Crenshaw}, {Kraemer}, \&
  {Schmitt}}]{fischer13}
{Fischer}, T.~C., {Crenshaw}, D.~M., {Kraemer}, S.~B., \& {Schmitt}, H.~R.
  2013, \apjs, 209, 1, arXiv:1308.4129

\bibitem[{{Gandhi} {et~al.}(2004){Gandhi}, {Crawford}, {Fabian}, \&
  {Johnstone}}]{g04}
{Gandhi}, P., {Crawford}, C.~S., {Fabian}, A.~C., \& {Johnstone}, R.~M. 2004,
  \mnras, 348, 529, astro-ph/0310772

\bibitem[{{Gandhi} \& {Fabian}(2003)}]{g03}
{Gandhi}, P., \& {Fabian}, A.~C. 2003, \mnras, 339, 1095, astro-ph/0211129

\bibitem[{{Gandhi} {et~al.}(2006){Gandhi}, {Fabian}, \&
  {Crawford}}]{g06_4c3929}
{Gandhi}, P., {Fabian}, A.~C., \& {Crawford}, C.~S. 2006, \mnras, 369, 1566,
  arXiv:astro-ph/0604184

\bibitem[{{Gandhi} {et~al.}(2007){Gandhi}, {Fabian}, {Suebsuwong}, {Malzac},
  {Miniutti}, \& {Wilman}}]{g07}
{Gandhi}, P., {Fabian}, A.~C., {Suebsuwong}, T., {Malzac}, J., {Miniutti}, G.,
  \& {Wilman}, R.~J. 2007, \mnras, 382, 1005, arXiv:0709.1984

\bibitem[{{Gandhi} {et~al.}(2009){Gandhi}, {Horst}, {Smette}, {H{\"o}nig},
  {Comastri}, {Gilli}, {Vignali}, \& {Duschl}}]{g09_mirxray}
{Gandhi}, P., {Horst}, H., {Smette}, A., {H{\"o}nig}, S., {Comastri}, A.,
  {Gilli}, R., {Vignali}, C., \& {Duschl}, W. 2009, \aap, 502, 457,
  arXiv:0902.2777

\bibitem[{{Gandhi} {et~al.}(2013){Gandhi}, {Terashima}, {Yamada}, {Mushotzky},
  {Ueda}, {Baumgartner}, {Alexander}, {Malzac}, {Vaghmare}, {Takahashi}, \&
  {Done}}]{g13_eso565}
{Gandhi}, P. {et~al.} 2013, \apj, 773, 51, arXiv:1305.4901

\bibitem[{{Georgantopoulos} {et~al.}(2013){Georgantopoulos}, {Comastri},
  {Vignali}, {Ranalli}, {Rovilos}, {Iwasawa}, {Gilli}, {Cappelluti}, {Carrera},
  {Fritz}, {Brusa}, {Elbaz}, {Mullaney}, {Castello-Mor}, {Barcons}, {Tozzi},
  {Balestra}, \& {Falocco}}]{georgantopoulos13}
{Georgantopoulos}, I. {et~al.} 2013, \aap, 555, A43

\bibitem[{{Gilli} {et~al.}(2007){Gilli}, {Comastri}, \& {Hasinger}}]{gilli07}
{Gilli}, R., {Comastri}, A., \& {Hasinger}, G. 2007, \aap, 463, 79,
  astro-ph/0610939

\bibitem[{{Gonz{\'a}lez Delgado} {et~al.}(2001){Gonz{\'a}lez Delgado},
  {Heckman}, \& {Leitherer}}]{gonzalezdelgado01}
{Gonz{\'a}lez Delgado}, R.~M., {Heckman}, T., \& {Leitherer}, C. 2001, \apj,
  546, 845, arXiv:astro-ph/0008417

\bibitem[{{Gonz{\'a}lez-Mart{\'{\i}}n}
  {et~al.}(2011){Gonz{\'a}lez-Mart{\'{\i}}n}, {Papadakis}, {Braito},
  {Masegosa}, {M{\'a}rquez}, {Mateos}, {Acosta-Pulido}, {Mart{\'{\i}}nez},
  {Ebrero}, {Esquej}, {O'Brien}, {Tueller}, {Warwick}, \&
  {Watson}}]{gonzalezmartin11}
{Gonz{\'a}lez-Mart{\'{\i}}n}, O. {et~al.} 2011, \aap, 527, A142,
  arXiv:1012.3080

\bibitem[{{Goulding} {et~al.}(2012){Goulding}, {Alexander}, {Bauer}, {Forman},
  {Hickox}, {Jones}, {Mullaney}, \& {Trichas}}]{goulding12}
{Goulding}, A.~D., {Alexander}, D.~M., {Bauer}, F.~E., {Forman}, W.~R.,
  {Hickox}, R.~C., {Jones}, C., {Mullaney}, J.~R., \& {Trichas}, M. 2012, \apj,
  755, 5, arXiv:1205.1800

\bibitem[{{Greenhill} {et~al.}(2008){Greenhill}, {Tilak}, \&
  {Madejski}}]{greenhill08}
{Greenhill}, L.~J., {Tilak}, A., \& {Madejski}, G. 2008, \apjl, 686, L13,
  0809.1108

\bibitem[{{Guainazzi} \& {Bianchi}(2007)}]{guainazzi07}
{Guainazzi}, M., \& {Bianchi}, S. 2007, \mnras, 374, 1290,
  arXiv:astro-ph/0610715

\bibitem[{{Guainazzi} {et~al.}(2009){Guainazzi}, {Risaliti}, {Nucita}, {Wang},
  {Bianchi}, {Soria}, \& {Zezas}}]{guainazzi09}
{Guainazzi}, M., {Risaliti}, G., {Nucita}, A., {Wang}, J., {Bianchi}, S.,
  {Soria}, R., \& {Zezas}, A. 2009, \aap, 505, 589, 0908.0268

\bibitem[{{Harrison} {et~al.}(2014){Harrison}, {Alexander}, {Mullaney}, \&
  {Swinbank}}]{harrison14}
{Harrison}, C.~M., {Alexander}, D.~M., {Mullaney}, J.~R., \& {Swinbank}, A.~M.
  2014, MNRAS in press, arXiv:1403.3086

\bibitem[{{Harrison} {et~al.}(2013){Harrison}, {Craig}, {Christensen},
  {Hailey}, {Zhang}, {Boggs}, {Stern}, {Cook}, {Forster}, {Giommi},
  {Grefenstette}, {Kim}, {Kitaguchi}, {Koglin}, {Madsen}, {Mao}, {Miyasaka},
  {Mori}, {Perri}, {Pivovaroff}, {Puccetti}, {Rana}, {Westergaard}, {Willis},
  {Zoglauer}, {An}, {Bachetti}, {Barri{\`e}re}, {Bellm}, {Bhalerao},
  {Brejnholt}, {Fuerst}, {Liebe}, {Markwardt}, {Nynka}, {Vogel}, {Walton},
  {Wik}, {Alexander}, {Cominsky}, {Hornschemeier}, {Hornstrup}, {Kaspi},
  {Madejski}, {Matt}, {Molendi}, {Smith}, {Tomsick}, {Ajello}, {Ballantyne},
  {Balokovi{\'c}}, {Barret}, {Bauer}, {Blandford}, {Brandt}, {Brenneman},
  {Chiang}, {Chakrabarty}, {Chenevez}, {Comastri}, {Dufour}, {Elvis}, {Fabian},
  {Farrah}, {Fryer}, {Gotthelf}, {Grindlay}, {Helfand}, {Krivonos}, {Meier},
  {Miller}, {Natalucci}, {Ogle}, {Ofek}, {Ptak}, {Reynolds}, {Rigby},
  {Tagliaferri}, {Thorsett}, {Treister}, \& {Urry}}]{nustar}
{Harrison}, F.~A. {et~al.} 2013, \apj, 770, 103, arXiv:1301.7307

\bibitem[{{Heckman} {et~al.}(1981){Heckman}, {Miley}, {van Breugel}, \&
  {Butcher}}]{heckman81}
{Heckman}, T.~M., {Miley}, G.~K., {van Breugel}, W.~J.~M., \& {Butcher}, H.~R.
  1981, \apj, 247, 403

\bibitem[{{Henkel} {et~al.}(2005){Henkel}, {Peck}, {Tarchi}, {Nagar}, {Braatz},
  {Castangia}, \& {Moscadelli}}]{henkel05}
{Henkel}, C., {Peck}, A.~B., {Tarchi}, A., {Nagar}, N.~M., {Braatz}, J.~A.,
  {Castangia}, P., \& {Moscadelli}, L. 2005, \aap, 436, 75, astro-ph/0503070

\bibitem[{{Hickox} {et~al.}(2014){Hickox}, {Mullaney}, {Alexander}, {Chen},
  {Civano}, {Goulding}, \& {Hainline}}]{hickox14}
{Hickox}, R.~C., {Mullaney}, J.~R., {Alexander}, D.~M., {Chen}, C.-T.~J.,
  {Civano}, F.~M., {Goulding}, A.~D., \& {Hainline}, K.~N. 2014, \apj, 782, 9,
  arXiv:1306.3218

\bibitem[{{H{\"o}nig} {et~al.}(2014){H{\"o}nig}, {Gandhi}, {Asmus},
  {Mushotzky}, {Antonucci}, {Ueda}, \& {Ichikawa}}]{hoenig14}
{H{\"o}nig}, S.~F., {Gandhi}, P., {Asmus}, D., {Mushotzky}, R.~F., {Antonucci},
  R., {Ueda}, Y., \& {Ichikawa}, K. 2014, \mnras, 438, 647, 1311.4880

\bibitem[{{Hopkins} \& {Elvis}(2010)}]{hopkinselvis10}
{Hopkins}, P.~F., \& {Elvis}, M. 2010, \mnras, 401, 7, arXiv:0904.0649

\bibitem[{{Jia} {et~al.}(2013){Jia}, {Ptak}, {Heckman}, \& {Zakamska}}]{jia13}
{Jia}, J., {Ptak}, A., {Heckman}, T., \& {Zakamska}, N.~L. 2013, \apj, 777, 27,
  arXiv:1205.0033

\bibitem[{{Kennicutt}(1998{\natexlab{a}})}]{kennicutt98}
{Kennicutt}, Jr., R.~C. 1998{\natexlab{a}}, \araa, 36, 189, astro-ph/9807187

\bibitem[{{Kennicutt}(1998{\natexlab{b}})}]{kennicutt98_ks}
------. 1998{\natexlab{b}}, \apj, 498, 541, astro-ph/9712213

\bibitem[{{Kirsch} {et~al.}(2004){Kirsch}, {Altieri}, {Chen}, {Haberl},
  {Metcalfe}, {Pollock}, {Read}, {Saxton}, {Sembay}, \& {Smith}}]{kirsch04}
{Kirsch}, M.~G.~F. {et~al.} 2004, in Society of Photo-Optical Instrumentation
  Engineers (SPIE) Conference Series, Vol. 5488, UV and Gamma-Ray Space
  Telescope Systems, ed. G.~{Hasinger} \& M.~J.~L. {Turner}, 103--114,
  astro-ph/0407257

\bibitem[{{Komossa} {et~al.}(2003){Komossa}, {Burwitz}, {Hasinger}, {Predehl},
  {Kaastra}, \& {Ikebe}}]{komossa03}
{Komossa}, S., {Burwitz}, V., {Hasinger}, G., {Predehl}, P., {Kaastra}, J.~S.,
  \& {Ikebe}, Y. 2003, \apjl, 582, L15, arXiv:astro-ph/0212099

\bibitem[{{Konami} {et~al.}(2011){Konami}, {Matsushita}, {Tsuru}, {Gandhi}, \&
  {Tamagawa}}]{konami11}
{Konami}, S., {Matsushita}, K., {Tsuru}, T.~G., {Gandhi}, P., \& {Tamagawa}, T.
  2011, \pasj, 63, 913, arXiv:1108.5778

\bibitem[{{Kondratko} {et~al.}(2006){Kondratko}, {Greenhill}, \&
  {Moran}}]{kondratko06b}
{Kondratko}, P.~T., {Greenhill}, L.~J., \& {Moran}, J.~M. 2006, \apj, 652, 136,
  astro-ph/0610060

\bibitem[{{Koss} {et~al.}(2013){Koss}, {Mushotzky}, {Baumgartner}, {Veilleux},
  {Tueller}, {Markwardt}, \& {Casey}}]{koss13}
{Koss}, M., {Mushotzky}, R., {Baumgartner}, W., {Veilleux}, S., {Tueller}, J.,
  {Markwardt}, C., \& {Casey}, C.~M. 2013, \apjl, 765, L26, 1302.0850

\bibitem[{{Kuo} {et~al.}(2011){Kuo}, {Braatz}, {Condon}, {Impellizzeri}, {Lo},
  {Zaw}, {Schenker}, {Henkel}, {Reid}, \& {Greene}}]{kuo11}
{Kuo}, C.~Y. {et~al.} 2011, \apj, 727, 20, 1008.2146

\bibitem[{{Lacy} {et~al.}(2004){Lacy}, {Storrie-Lombardi}, {Sajina},
  {Appleton}, {Armus}, {Chapman}, {Choi}, {Fadda}, {Fang}, {Frayer},
  {Heinrichsen}, {Helou}, {Im}, {Marleau}, {Masci}, {Shupe}, {Soifer},
  {Surace}, {Teplitz}, {Wilson}, \& {Yan}}]{lacy04}
{Lacy}, M. {et~al.} 2004, \apjs, 154, 166

\bibitem[{{LaMassa et al.}(2014)}]{lamassa14}
{LaMassa et al.} 2014, ApJ in press, arXiv:1404.0012L

\bibitem[{{Lansbury} {et~al.}(2014){Lansbury}, {Alexander}, {Del Moro},
  {Gandhi}, {Assef}, {Stern}, {Aird}, {Ballantyne}, {Balokovi{\'c}}, {Bauer},
  {Boggs}, {Brandt}, {Christensen}, {Craig}, {Elvis}, {Grefenstette}, {Hailey},
  {Harrison}, {Hickox}, {Koss}, {LaMassa}, {Luo}, {Mullaney}, {Teng}, {Urry},
  \& {Zhang}}]{lansbury14}
{Lansbury}, G.~B. {et~al.} 2014, \apj, 785, 17, arXiv:1402.2666

\bibitem[{{Liu} {et~al.}(2013){Liu}, {Zakamska}, {Greene}, {Nesvadba}, \&
  {Liu}}]{liu13}
{Liu}, G., {Zakamska}, N.~L., {Greene}, J.~E., {Nesvadba}, N.~P.~H., \& {Liu},
  X. 2013, \mnras, 436, 2576, 1305.6922

\bibitem[{{Magdziarz} \& {Zdziarski}(1995)}]{pexrav}
{Magdziarz}, P., \& {Zdziarski}, A.~A. 1995, \mnras, 273, 837

\bibitem[{{Mainieri} {et~al.}(2011){Mainieri}, {Bongiorno}, {Merloni}, {Aller},
  {Carollo}, {Iwasawa}, {Koekemoer}, {Mignoli}, {Silverman}, {Bolzonella},
  {Brusa}, {Comastri}, {Gilli}, {Halliday}, {Ilbert}, {Lusso}, {Salvato},
  {Vignali}, {Zamorani}, {Contini}, {Kneib}, {Le F{\`e}vre}, {Lilly},
  {Renzini}, {Scodeggio}, {Balestra}, {Bardelli}, {Caputi}, {Coppa},
  {Cucciati}, {de la Torre}, {de Ravel}, {Franzetti}, {Garilli}, {Iovino},
  {Kampczyk}, {Knobel}, {Kova{\v c}}, {Lamareille}, {Le Borgne}, {Le Brun},
  {Maier}, {Nair}, {Pello}, {Peng}, {Perez Montero}, {Pozzetti},
  {Ricciardelli}, {Tanaka}, {Tasca}, {Tresse}, {Vergani}, {Zucca}, {Aussel},
  {Capak}, {Cappelluti}, {Elvis}, {Fiore}, {Hasinger}, {Impey}, {Le Floc'h},
  {Scoville}, {Taniguchi}, \& {Trump}}]{mainieri11}
{Mainieri}, V. {et~al.} 2011, \aap, 535, A80, arXiv:1105.5395

\bibitem[{{Marconi} \& {Hunt}(2003)}]{marconihunt03}
{Marconi}, A., \& {Hunt}, L.~K. 2003, \apjl, 589, L21, astro-ph/0304274

\bibitem[{{Mateos} {et~al.}(2012){Mateos}, {Alonso-Herrero}, {Carrera},
  {Blain}, {Watson}, {Barcons}, {Braito}, {Severgnini}, {Donley}, \&
  {Stern}}]{mateos12}
{Mateos}, S. {et~al.} 2012, \mnras, 426, 3271, arXiv:1208.2530

\bibitem[{{Mateos} {et~al.}(2005){Mateos}, {Barcons}, {Carrera}, {Ceballos},
  {Caccianiga}, {Lamer}, {Maccacaro}, {Page}, {Schwope}, \&
  {Watson}}]{mateos05_wide}
------. 2005, \aap, 433, 855

\bibitem[{{Matt} {et~al.}(2000){Matt}, {Fabian}, {Guainazzi}, {Iwasawa},
  {Bassani}, \& {Malaguti}}]{matt00}
{Matt}, G., {Fabian}, A.~C., {Guainazzi}, M., {Iwasawa}, K., {Bassani}, L., \&
  {Malaguti}, G. 2000, \mnras, 318, 173

\bibitem[{{McCarthy}(1993)}]{mccarthy93}
{McCarthy}, P.~J. 1993, \araa, 31, 639

\bibitem[{{McConnell} \& {Ma}(2013)}]{mcconnellma13}
{McConnell}, N.~J., \& {Ma}, C.-P. 2013, \apj, 764, 184, arXiv:1211.2816

\bibitem[{{Merloni} {et~al.}(2014){Merloni}, {Bongiorno}, {Brusa}, {Iwasawa},
  {Mainieri}, {Magnelli}, {Salvato}, {Berta}, {Cappelluti}, {Comastri},
  {Fiore}, {Gilli}, {Koekemoer}, {Le Floc'h}, {Lusso}, {Lutz}, {Miyaji},
  {Pozzi}, {Riguccini}, {Rosario}, {Silverman}, {Symeonidis}, {Treister},
  {Vignali}, \& {Zamorani}}]{merloni14}
{Merloni}, A. {et~al.} 2014, \mnras, 437, 3550, arXiv:1311.1305

\bibitem[{{Miley} \& {De Breuck}(2008)}]{mileydebreuck08}
{Miley}, G., \& {De Breuck}, C. 2008, \aapr, 15, 67, 0802.2770

\bibitem[{{Mineo} {et~al.}(2012){Mineo}, {Gilfanov}, \&
  {Sunyaev}}]{mineo12_hotgas}
{Mineo}, S., {Gilfanov}, M., \& {Sunyaev}, R. 2012, \mnras, 426, 1870,
  arXiv:1205.3715

\bibitem[{{Miniutti} {et~al.}(2014){Miniutti}, {Sanfrutos}, {Beuchert},
  {Ag{\'{\i}}s-Gonz{\'a}lez}, {Longinotti}, {Piconcelli}, {Krongold},
  {Guainazzi}, {Bianchi}, {Matt}, \& {Jim{\'e}nez-Bail{\'o}n}}]{miniutti14}
{Miniutti}, G. {et~al.} 2014, \mnras, 437, 1776, 1310.7701

\bibitem[{{Mulchaey} {et~al.}(1994){Mulchaey}, {Koratkar}, {Ward}, {Wilson},
  {Whittle}, {Antonucci}, {Kinney}, \& {Hurt}}]{mulchaey94}
{Mulchaey}, J.~S., {Koratkar}, A., {Ward}, M.~J., {Wilson}, A.~S., {Whittle},
  M., {Antonucci}, R.~R.~J., {Kinney}, A.~L., \& {Hurt}, T. 1994, \apj, 436,
  586

\bibitem[{{Mullaney} {et~al.}(2013){Mullaney}, {Alexander}, {Fine}, {Goulding},
  {Harrison}, \& {Hickox}}]{mullaney13}
{Mullaney}, J.~R., {Alexander}, D.~M., {Fine}, S., {Goulding}, A.~D.,
  {Harrison}, C.~M., \& {Hickox}, R.~C. 2013, \mnras, 433, 622, 1305.0263

\bibitem[{{Murphy} \& {Yaqoob}(2009)}]{mytorus}
{Murphy}, K.~D., \& {Yaqoob}, T. 2009, \mnras, 397, 1549, arXiv:0905.3188

\bibitem[{{Nair} \& {Abraham}(2010)}]{nair10}
{Nair}, P.~B., \& {Abraham}, R.~G. 2010, \apjs, 186, 427, 1001.2401

\bibitem[{{Nandra} {et~al.}(2013){Nandra}, {Barret}, {Barcons}, {Fabian}, {den
  Herder}, {Piro}, {Watson}, {Adami}, {Aird}, {Afonso}, \& et~al.}]{athenaplus}
{Nandra}, K. {et~al.} 2013, The Hot and Energetic Universe: A White Paper
  presenting the science theme motivating the Athena+ mission, arXiv:1306.2307

\bibitem[{{Nandra} {et~al.}(1997){Nandra}, {George}, {Mushotzky}, {Turner}, \&
  {Yaqoob}}]{nandra97}
{Nandra}, K., {George}, I.~M., {Mushotzky}, R.~F., {Turner}, T.~J., \&
  {Yaqoob}, T. 1997, \apjl, 488, L91+, arXiv:astro-ph/9708030

\bibitem[{{Nandra} {et~al.}(2007){Nandra}, {O'Neill}, {George}, \&
  {Reeves}}]{pexmon}
{Nandra}, K., {O'Neill}, P.~M., {George}, I.~M., \& {Reeves}, J.~N. 2007,
  \mnras, 382, 194, 0708.1305

\bibitem[{{Netzer} {et~al.}(2006){Netzer}, {Mainieri}, {Rosati}, \&
  {Trakhtenbrot}}]{netzer06}
{Netzer}, H., {Mainieri}, V., {Rosati}, P., \& {Trakhtenbrot}, B. 2006, \aap,
  453, 525, astro-ph/0603712

\bibitem[{{Norman} {et~al.}(2002){Norman}, {Hasinger}, {Giacconi}, {Gilli},
  {Kewley}, {Nonino}, {Rosati}, {Szokoly}, {Tozzi}, {Wang}, {Zheng}, {Zirm},
  {Bergeron}, {Gilmozzi}, {Grogin}, {Koekemoer}, \& {Schreier}}]{norman02}
{Norman}, C. {et~al.} 2002, \apj, 571, 218

\bibitem[{{Oh} {et~al.}(2011){Oh}, {Sarzi}, {Schawinski}, \& {Yi}}]{oh11}
{Oh}, K., {Sarzi}, M., {Schawinski}, K., \& {Yi}, S.~K. 2011, \apjs, 195, 13,
  arXiv:1106.1896

\bibitem[{{Panessa} {et~al.}(2006){Panessa}, {Bassani}, {Cappi}, {Dadina},
  {Barcons}, {Carrera}, {Ho}, \& {Iwasawa}}]{panessa06}
{Panessa}, F., {Bassani}, L., {Cappi}, M., {Dadina}, M., {Barcons}, X.,
  {Carrera}, F.~J., {Ho}, L.~C., \& {Iwasawa}, K. 2006, \aap, 455, 173,
  arXiv:astro-ph/0605236

\bibitem[{{Pfefferkorn} {et~al.}(2001){Pfefferkorn}, {Boller}, \&
  {Rafanelli}}]{pfefferkorn01}
{Pfefferkorn}, F., {Boller}, T., \& {Rafanelli}, P. 2001, \aap, 368, 797,
  astro-ph/0101184

\bibitem[{{Piconcelli} {et~al.}(2005){Piconcelli}, {Jimenez-Bail{\'o}n},
  {Guainazzi}, {Schartel}, {Rodr{\'{\i}}guez-Pascual}, \&
  {Santos-Lle{\'o}}}]{piconcelli05}
{Piconcelli}, E., {Jimenez-Bail{\'o}n}, E., {Guainazzi}, M., {Schartel}, N.,
  {Rodr{\'{\i}}guez-Pascual}, P.~M., \& {Santos-Lle{\'o}}, M. 2005, \aap, 432,
  15, arXiv:astro-ph/0411051

\bibitem[{{Planck Collaboration}(2013)}]{planckcosmology}
{Planck Collaboration}. 2013, A\&A submitted, arXiv:1303.5076

\bibitem[{{Ptak} {et~al.}(2006){Ptak}, {Zakamska}, {Strauss}, {Krolik},
  {Heckman}, {Schneider}, \& {Brinkmann}}]{ptak06}
{Ptak}, A., {Zakamska}, N.~L., {Strauss}, M.~A., {Krolik}, J.~H., {Heckman},
  T.~M., {Schneider}, D.~P., \& {Brinkmann}, J. 2006, \apj, 637, 147,
  astro-ph/0510204

\bibitem[{{Ptak et al.}(2014)}]{ptak14}
{Ptak et al.} 2014, ApJ submitted

\bibitem[{{Puccetti et al.}(2014)}]{puccetti14}
{Puccetti et al.} 2014, ApJ submitted

\bibitem[{{Reyes} {et~al.}(2008){Reyes}, {Zakamska}, {Strauss}, {Green},
  {Krolik}, {Shen}, {Richards}, {Anderson}, \& {Schneider}}]{reyes08}
{Reyes}, R. {et~al.} 2008, \aj, 136, 2373, arXiv:0801.1115

\bibitem[{{Ricci} {et~al.}(2011){Ricci}, {Walter}, {Courvoisier}, \&
  {Paltani}}]{ricci11}
{Ricci}, C., {Walter}, R., {Courvoisier}, T.~J.-L., \& {Paltani}, S. 2011,
  \aap, 532, A102, arXiv:1104.3676

\bibitem[{{Risaliti} {et~al.}(2005){Risaliti}, {Elvis}, {Fabbiano}, {Baldi}, \&
  {Zezas}}]{risaliti05}
{Risaliti}, G., {Elvis}, M., {Fabbiano}, G., {Baldi}, A., \& {Zezas}, A. 2005,
  \apjl, 623, L93

\bibitem[{{Sanders} {et~al.}(2003){Sanders}, {Mazzarella}, {Kim}, {Surace}, \&
  {Soifer}}]{sanders03_rbgs}
{Sanders}, D.~B., {Mazzarella}, J.~M., {Kim}, D.-C., {Surace}, J.~A., \&
  {Soifer}, B.~T. 2003, \aj, 126, 1607, astro-ph/0306263

\bibitem[{{Schlafly} \& {Finkbeiner}(2011)}]{schlafly11}
{Schlafly}, E.~F., \& {Finkbeiner}, D.~P. 2011, \apj, 737, 103, arXiv:1012.4804

\bibitem[{{Severgnini} {et~al.}(2011){Severgnini}, {Caccianiga}, {Della Ceca},
  {Braito}, {Vignali}, {La Parola}, \& {Moretti}}]{severgnini11}
{Severgnini}, P., {Caccianiga}, A., {Della Ceca}, R., {Braito}, V., {Vignali},
  C., {La Parola}, V., \& {Moretti}, A. 2011, \aap, 525, A38, 1010.2085

\bibitem[{{Smith} {et~al.}(2001){Smith}, {Brickhouse}, {Liedahl}, \&
  {Raymond}}]{apec}
{Smith}, R.~K., {Brickhouse}, N.~S., {Liedahl}, D.~A., \& {Raymond}, J.~C.
  2001, \apjl, 556, L91, arXiv:astro-ph/0106478

\bibitem[{{Stern} {et~al.}(2012){Stern}, {Assef}, {Benford}, {Blain}, {Cutri},
  {Dey}, {Eisenhardt}, {Griffith}, {Jarrett}, {Lake}, {Masci}, {Petty},
  {Stanford}, {Tsai}, {Wright}, {Yan}, {Harrison}, \& {Madsen}}]{stern12}
{Stern}, D. {et~al.} 2012, \apj, 753, 30, 1205.0811

\bibitem[{{Stern} {et~al.}(2005){Stern}, {Eisenhardt}, {Gorjian}, {Kochanek},
  {Caldwell}, {Eisenstein}, {Brodwin}, {Brown}, {Cool}, {Dey}, {Green},
  {Jannuzi}, {Murray}, {Pahre}, \& {Willner}}]{stern05}
------. 2005, \apj, 631, 163, astro-ph/0410523

\bibitem[{{Stern} {et~al.}(2014){Stern}, {Lansbury}, {Assef}, {Brandt},
  {Alexander}, {Ballantyne}, {Balokovic}, {Benford}, {Blain}, {Boggs},
  {Bridge}, {Brightman}, {Christensen}, {Comastri}, {Craig}, {Del Moro},
  {Eisenhardt}, {Gandhi}, {Griffith}, {Hailey}, {Harrison}, {Hickox},
  {Jarrett}, {Koss}, {Lake}, {LaMassa}, {Luo}, {Tsai}, {Walton}, {Wright},
  {Wu}, {Yan}, \& {Zhang}}]{stern14}
------. 2014, ApJ submitted, arXiv:1403.3078

\bibitem[{{Stern} {et~al.}(2002){Stern}, {Moran}, {Coil}, {Connolly}, {Davis},
  {Dawson}, {Dey}, {Eisenhardt}, {Elston}, {Graham}, {Harrison}, {Helfand},
  {Holden}, {Mao}, {Rosati}, {Spinrad}, {Stanford}, {Tozzi}, \& {Wu}}]{stern02}
------. 2002, \apj, 568, 71

\bibitem[{{Stoklasov{\'a}} {et~al.}(2009){Stoklasov{\'a}}, {Ferruit},
  {Emsellem}, {Jungwiert}, {P{\'e}contal}, \& {S{\'a}nchez}}]{stoklasova09}
{Stoklasov{\'a}}, I., {Ferruit}, P., {Emsellem}, E., {Jungwiert}, B.,
  {P{\'e}contal}, E., \& {S{\'a}nchez}, S.~F. 2009, \aap, 500, 1287, 0905.3349

\bibitem[{{Takahashi} {et~al.}(2012){Takahashi}, {Mitsuda}, {Kelley}, {Aarts},
  {Aharonian}, {Akamatsu}, {Akimoto}, {Allen}, {Anabuki}, {Angelini}, {Arnaud},
  {Asai}, {Audard}, {Awaki}, {Azzarello}, {Baluta}, {Bamba}, {Bando}, {Bautz},
  {Blandford}, {Boyce}, {Brown}, {Cackett}, {Chernyakova}, {Coppi},
  {Costantini}, {de Plaa}, {den Herder}, {DiPirro}, {Done}, {Dotani}, {Doty},
  {Ebisawa}, {Eckart}, {Enoto}, {Ezoe}, {Fabian}, {Ferrigno}, {Foster},
  {Fujimoto}, {Fukazawa}, {Funk}, {Furuzawa}, {Galeazzi}, {Gallo}, {Gandhi},
  {Gendreau}, {Gilmore}, {Haas}, {Haba}, {Hamaguchi}, {Hatsukade}, {Hayashi},
  {Hayashida}, {Hiraga}, {Hirose}, {Hornschemeier}, {Hoshino}, {Hughes},
  {Hwang}, {Iizuka}, {Inoue}, {Ishibashi}, {Ishida}, {Ishimura}, {Ishisaki},
  {Ito}, {Iwata}, {Iyomoto}, {Kaastra}, {Kallman}, {Kamae}, {Kataoka},
  {Katsuda}, {Kawahara}, {Kawaharada}, {Kawai}, {Kawasaki}, {Khangaluyan},
  {Kilbourne}, {Kimura}, {Kinugasa}, {Kitamoto}, {Kitayama}, {Kohmura},
  {Kokubun}, {Kosaka}, {Koujelev}, {Koyama}, {Krimm}, {Kubota}, {Kunieda},
  {LaMassa}, {Laurent}, {Lebrun}, {Leutenegger}, {Limousin}, {Loewenstein},
  {Long}, {Lumb}, {Madejski}, {Maeda}, {Makishima}, {Marchand}, {Markevitch},
  {Matsumoto}, {Matsushita}, {McCammon}, {McNamara}, {Miller}, {Miller},
  {Mineshige}, {Minesugi}, {Mitsuishi}, {Miyazawa}, {Mizuno}, {Mori}, {Mori},
  {Mukai}, {Murakami}, {Murakami}, {Mushotzky}, {Nagano}, {Nagino}, {Nakagawa},
  {Nakajima}, {Nakamori}, {Nakazawa}, {Namba}, {Natsukari}, {Nishioka},
  {Nobukawa}, {Nomachi}, {O'Dell}, {Odaka}, {Ogawa}, {Ogawa}, {Ogi}, {Ohashi},
  {Ohno}, {Ohta}, {Okajima}, {Okamoto}, {Okazaki}, {Ota}, {Ozaki}, {Paerels},
  {Paltani}, {Parmar}, {Petre}, {Pohl}, {Porter}, {Ramsey}, {Reis}, {Reynolds},
  {Russell}, {Safi-Harb}, {Sakai}, {Sameshima}, {Sanders}, {Sato}, {Sato},
  {Sato}, {Sato}, {Sawada}, {Serlemitsos}, {Seta}, {Shibano}, {Shida},
  {Shimada}, {Shinozaki}, {Shirron}, {Simionescu}, {Simmons}, {Smith},
  {Sneiderman}, {Soong}, {Stawarz}, {Sugawara}, {Sugita}, {Sugita},
  {Szymkowiak}, {Tajima}, {Takahashi}, {Takeda}, {Takei}, {Tamagawa}, {Tamura},
  {Tamura}, {Tanaka}, {Tanaka}, {Tashiro}, {Tawara}, {Terada}, {Terashima},
  {Tombesi}, {Tomida}, {Tsuboi}, {Tsujimoto}, {Tsunemi}, {Tsuru}, {Uchida},
  {Uchiyama}, {Uchiyama}, {Ueda}, {Ueno}, {Uno}, {Urry}, {Ursino}, {de Vries},
  {Wada}, {Watanabe}, {Werner}, {White}, {Yamada}, {Yamada}, {Yamaguchi},
  {Yamasaki}, {Yamauchi}, {Yamauchi}, {Yatsu}, {Yonetoku}, {Yoshida}, \&
  {Yuasa}}]{astroh12}
{Takahashi}, T. {et~al.} 2012, in Society of Photo-Optical Instrumentation
  Engineers (SPIE) Conference Series, Vol. 8443, Society of Photo-Optical
  Instrumentation Engineers (SPIE) Conference Series, arXiv:1210.4378

\bibitem[{{Teng} {et~al.}(2014){Teng}, {Brandt}, {Harrison}, {Luo},
  {Alexander}, {Bauer}, {Boggs}, {Christensen}, {Comastri}, {Craig}, {Fabian},
  {Farrah}, {Fiore}, {Gandhi}, {Grefenstette}, {Hailey}, {Hickox}, {Madsen},
  {Ptak}, {Rigby}, {Risaliti}, {Saez}, {Stern}, {Veilleux}, {Walton}, {Wik}, \&
  {Zhang}}]{teng14}
{Teng}, S.~H. {et~al.} 2014, \apj, 785, 19, arXiv:1402.4811

\bibitem[{{Tozzi} {et~al.}(2009){Tozzi}, {Mainieri}, {Rosati}, {Padovani},
  {Kellermann}, {Fomalont}, {Miller}, {Shaver}, {Bergeron}, {Brandt}, {Brusa},
  {Giacconi}, {Hasinger}, {Lehmer}, {Nonino}, {Norman}, \&
  {Silverman}}]{tozzi09}
{Tozzi}, P. {et~al.} 2009, \apj, 698, 740, 0902.2930

\bibitem[{{Treister} {et~al.}(2009){Treister}, {Urry}, \&
  {Virani}}]{treister09}
{Treister}, E., {Urry}, C.~M., \& {Virani}, S. 2009, \apj, 696, 110,
  arXiv:0902.0608

\bibitem[{{Tremaine} {et~al.}(2002){Tremaine}, {Gebhardt}, {Bender}, {Bower},
  {Dressler}, {Faber}, {Filippenko}, {Green}, {Grillmair}, {Ho}, {Kormendy},
  {Lauer}, {Magorrian}, {Pinkney}, \& {Richstone}}]{tremaine02}
{Tremaine}, S. {et~al.} 2002, \apj, 574, 740, astro-ph/0203468

\bibitem[{{Ueda} {et~al.}(2014){Ueda}, {Akiyama}, {Hasinger}, {Miyaji}, \&
  {Watson}}]{ueda14}
{Ueda}, Y., {Akiyama}, M., {Hasinger}, G., {Miyaji}, T., \& {Watson}, M.~G.
  2014, ApJ submitted, arXiv:1402.1836

\bibitem[{{Ueda} {et~al.}(2007){Ueda}, {Eguchi}, {Terashima}, {Mushotzky},
  {Tueller}, {Markwardt}, {Gehrels}, {Hashimoto}, \& {Potter}}]{ueda07}
{Ueda}, Y. {et~al.} 2007, \apjl, 664, L79, arXiv:0706.1168

\bibitem[{{Ulvestad} \& {Wilson}(1984)}]{ulvestad84_v}
{Ulvestad}, J.~S., \& {Wilson}, A.~S. 1984, \apj, 278, 544

\bibitem[{{Vasudevan} {et~al.}(2010){Vasudevan}, {Fabian}, {Gandhi}, {Winter},
  \& {Mushotzky}}]{vasudevan10}
{Vasudevan}, R.~V., {Fabian}, A.~C., {Gandhi}, P., {Winter}, L.~M., \&
  {Mushotzky}, R.~F. 2010, \mnras, 402, 1081, arXiv:0910.5256

\bibitem[{{Vasudevan} {et~al.}(2013){Vasudevan}, {Mushotzky}, \&
  {Gandhi}}]{vasudevan13_localxrb}
{Vasudevan}, R.~V., {Mushotzky}, R.~F., \& {Gandhi}, P. 2013, \apjl, 770, L37,
  arXiv:1305.6611

\bibitem[{{Veilleux} {et~al.}(2009){Veilleux}, {Rupke}, {Kim}, {Genzel},
  {Sturm}, {Lutz}, {Contursi}, {Schweitzer}, {Tacconi}, {Netzer}, {Sternberg},
  {Mihos}, {Baker}, {Mazzarella}, {Lord}, {Sanders}, {Stockton}, {Joseph}, \&
  {Barnes}}]{veilleux09}
{Veilleux}, S. {et~al.} 2009, \apjs, 182, 628, 0905.1577

\bibitem[{{Vignali} {et~al.}(2004){Vignali}, {Alexander}, \&
  {Comastri}}]{vignali04}
{Vignali}, C., {Alexander}, D.~M., \& {Comastri}, A. 2004, \mnras, 354, 720,
  astro-ph/0407293

\bibitem[{{Vignali} {et~al.}(2006){Vignali}, {Alexander}, \&
  {Comastri}}]{vignali06}
------. 2006, \mnras, 373, 321, astro-ph/0609089

\bibitem[{{Vignali} {et~al.}(2010){Vignali}, {Alexander}, {Gilli}, \&
  {Pozzi}}]{vignali10}
{Vignali}, C., {Alexander}, D.~M., {Gilli}, R., \& {Pozzi}, F. 2010, \mnras,
  404, 48, arXiv:1001.2005

\bibitem[{{Vignati} {et~al.}(1999){Vignati}, {Molendi}, {Matt}, {Guainazzi},
  {Antonelli}, {Bassani}, {Brandt}, {Fabian}, {Iwasawa}, {Maiolino},
  {Malaguti}, {Marconi}, \& {Perola}}]{vignati99}
{Vignati}, P. {et~al.} 1999, \aap, 349, L57, astro-ph/9908253

\bibitem[{{Wang} {et~al.}(2011){Wang}, {Fabbiano}, {Elvis}, {Risaliti},
  {Karovska}, {Zezas}, {Mundell}, {Dumas}, \& {Schinnerer}}]{wang11_ngc4151}
{Wang}, J. {et~al.} 2011, \apj, 742, 23, 1103.1913

\bibitem[{{Wang} {et~al.}(2014){Wang}, {Nardini}, {Fabbiano}, {Karovska},
  {Elvis}, {Pellegrini}, {Max}, {Risaliti}, {U}, \& {Zezas}}]{wang14_n6240}
------. 2014, \apj, 781, 55, 1303.2980

\bibitem[{{Wang} {et~al.}(2007){Wang}, {Chen}, {Yan}, {Hu}, \&
  {Bian}}]{wang07_feedback}
{Wang}, J.-M., {Chen}, Y.-M., {Yan}, C.-S., {Hu}, C., \& {Bian}, W.-H. 2007,
  \apjl, 661, L143, 0704.2288

\bibitem[{{Wilkes} {et~al.}(2013){Wilkes}, {Kuraszkiewicz}, {Haas}, {Barthel},
  {Leipski}, {Willner}, {Worrall}, {Birkinshaw}, {Antonucci}, {Ashby}, {Chini},
  {Fazio}, {Lawrence}, {Ogle}, \& {Schulz}}]{wilkes13}
{Wilkes}, B.~J. {et~al.} 2013, \apj, 773, 15, 1306.0850

\bibitem[{{Wright} {et~al.}(2010){Wright}, {Eisenhardt}, {Mainzer}, {Ressler},
  {Cutri}, {Jarrett}, {Kirkpatrick}, {Padgett}, {McMillan}, {Skrutskie},
  {Stanford}, {Cohen}, {Walker}, {Mather}, {Leisawitz}, {Gautier}, {McLean},
  {Benford}, {Lonsdale}, {Blain}, {Mendez}, {Irace}, {Duval}, {Liu}, {Royer},
  {Heinrichsen}, {Howard}, {Shannon}, {Kendall}, {Walsh}, {Larsen}, {Cardon},
  {Schick}, {Schwalm}, {Abid}, {Fabinsky}, {Naes}, \& {Tsai}}]{wise}
{Wright}, E.~L. {et~al.} 2010, \aj, 140, 1868, arXiv:1008.0031

\bibitem[{{Yaqoob}(2012)}]{yaqoob12}
{Yaqoob}, T. 2012, \mnras, 423, 3360, arXiv:1204.4196

\bibitem[{{Zakamska} {et~al.}(2003){Zakamska}, {Strauss}, {Krolik}, {Collinge},
  {Hall}, {Hao}, {Heckman}, {Ivezi{\'c}}, {Richards}, {Schlegel}, {Schneider},
  {Strateva}, {Vanden Berk}, {Anderson}, \& {Brinkmann}}]{zakamska03}
{Zakamska}, N.~L. {et~al.} 2003, \aj, 126, 2125, astro-ph/0309551

\end{thebibliography}
\end{document}